\documentclass[letterpaper,final,twocolumn]{IEEEtran}

\usepackage[T1]{fontenc}
\usepackage{amssymb,amsmath,epsfig}
\usepackage[latin5]{inputenc}
\usepackage{graphicx,times,latexsym}
\usepackage{subfigure}
\usepackage{psfrag,cite}
\usepackage{setspace}
\usepackage{algorithmic, algorithm}
\usepackage{color}
\usepackage{url}

\newtheorem{theorem}{Theorem}
\newtheorem{lemma}{Lemma}
\newtheorem{remark}{Remark}
\newtheorem{assume}{Assumption}
\newtheorem{defn}{Definition}

\newtheorem{Proposition}{Proposition}
\newcommand{\oprocendsymbol}{\hbox{$\bullet$}}
\newcommand{\oprocend}{\relax\ifmmode\else\unskip\hfill\fi\oprocendsymbol}

\newcommand{\blue}[1]{\color{black}{#1}}

\definecolor{dark-green}{rgb}{0,.4,0}

\newcommand{\magenta}[1]{{\color{magenta} #1}}
 \renewcommand{\magenta}[1]{{#1}}
\newcommand{\modulo}{\text{mod}}

\usepackage{mathtools}
\mathtoolsset{showonlyrefs=true}

\newcommand{\real}{\mathbb{R}}
\newcommand{\integers}{\mathbb{N}}
\newcommand{\complex}{\mathbb{C}}

\DeclareMathOperator*{\Tr}{Tr}
\DeclareMathOperator*{\RE}{Re}
\DeclareMathOperator*{\IM}{Im}
\newcommand{\floor}[1]{\lfloor #1 \rfloor}
\newcommand{\ceil}[1]{\lceil #1 \rceil}

\providecommand{\figref}{Fig.\,\ref}

\allowdisplaybreaks

\graphicspath{{epsfiles/}}

\title{Exploiting timing information in event-triggered stabilization of linear systems with disturbances 
}

\author{Mohammad Javad Khojasteh, Mojtaba Hedayatpour, Jorge
  Cort{\'e}s, Massimo Franceschetti \thanks{M.~J.~Khojasteh is with
    the Center for Autonomous Systems and Technologies, California Institute of
    Technology. M. Hedayatpour is with the Faculty of Engineering and Applied Science, University of Regina,
    Canada.  J. Cort{\'e}s is with the Department of Mechanical and
    Aerospace Engineering, University of California, San
    Diego. M. Franceschetti is with the Department of Electrical and
    Computer Engineering of University of California, San Diego.
    (e-mails: \texttt{mjkhojas@caltech.edu
      \{cortes,massimo\}@ucsd.edu, hedayatm@uregina.ca }).  This
    research was partially supported by NSF awards CNS-1446891 and ECCS-1917177.}}
 
\begin{document}
\maketitle
\begin{abstract}
  In the same way that subsequent pauses in spoken language are used
  to convey information, it is also possible to transmit information
  in communication \magenta{networks} not only by message content, but
  also with its timing.  This paper presents an event-triggering
  strategy that utilizes timing information by transmitting in a
  state-dependent fashion. We consider the stabilization of a
  continuous-time, time-invariant, linear plant over a digital
  communication channel with bounded delay and subject to bounded
  plant disturbances and establish two main results. On the one hand,
  we design an encoding-decoding scheme that guarantees a sufficient
  information transmission rate for stabilization. On the other hand,
  we determine a lower bound on the information transmission rate
  necessary for stabilization by any control policy.
\end{abstract}

\vspace*{-2ex}

\section{Introduction}\label{sec:intro}
\magenta{In many networked control systems} (NCS),
the feedback loop is closed over a communication channel\magenta{~\cite{hespanha2007survey}}.
In this context, data-rate theorems~\cite{Nair,fang2017towards} state that the minimum
communication rate to achieve stabilization is equal to the
\emph{entropy rate} of the plant, expressed by the sum of the unstable
modes in nats (one nat corresponds to $1/\ln 2$ bits.)
Key contributions by~\cite{Mitter}, \cite{nair2004stabilizability}, and~\cite{liberzon2003stabilization}
consider a ``bit-pipe" communication channel, capable of noiseless
transmission of a finite number of bits per unit time evolution of the plant.  Extensions to noisy communication channels are considered
in~\cite{sahai2006necessity,matveev2009estimation,Yukselbook,como2010anytime}.
Stabilization over time-varying bit-pipe channels, including the erasure channel as a special case, are studied in~\cite{Paolo}. 
Additional formulations include 
stabilization of switched linear 
systems~\cite{yang2017feedback}, uncertain systems~\cite{martins2006feedback}, nonlinear systems~\cite{de2005n,Topological},
multiplicative noise~\cite{ding2016multiplicative}, and optimal
control~\cite{kostina2016ratemm,toli}.

\magenta{
While the majority of communication networks transmit information by adjusting the content of the message, it is also possible to communicate information by adjusting the transmission time of a
symbol~\cite{anantharam1996bits}. 
In fact, it is known that event-triggering control techniques\cite{WPMHH-KHJ-PT:12} encode information in the timing in a state-dependent fashion~\cite{khojasteh2018stabilizing}.
The works ~\cite{tanwani2017stabilization,1807103d3,7562376,li2012stabilizingwwdw,tallapragada2016event,heemels2012periodic,tanwani2016observer} study event-triggered strategies over communication networks without exploiting the implicit timing information in the triggering events. In particular,~\cite{tallapragada2016event} studies the stabilization of the linear systems over finite data-rate channels with bounded delay.
The work~\cite{heemels2012periodic} considers periodic event-triggered control for linear systems where the event-triggering condition is verified periodically. 
The work~\cite{tanwani2016observer} considers output feedback stabilization of linear systems with no disturbance where the measured outputs and control inputs are subject to event-triggered sampling and dynamic quantization. 
}
%

\magenta{In contrast to the above works, to decrease the number of physical packets transmitted over
  the network (data payload), the
  works~\cite{guo2019optimal,pearson2017control,Level,ling2016bit,ling2017bit,linsenmayer2017delay,OurJournal1}
  study event-triggered strategies that exploit the inherent timing
  information in the events, and show that stability can be
  achieved with a rate lower than the one prescribed by data-rate
  theorems. The work \cite{guo2019optimal} utilizes the implicit timing
  information in triggering events to estimate a Wiener Process over a
  finite rate communication channel subject to finite
  delay. The work~\cite{pearson2017control} uses event-triggering to encode
  information in timing for stabilization of linear systems without
  disturbances in a silence-based communication
  manner~\cite{dhulipala2009silence}. Also,~\cite{yildiz2019event}
  extends the results of~\cite{pearson2017control} to optimal control.  The works~\cite{Level,ling2016bit} show that, with
  sufficiently small delays,
and assuming the controller has knowledge of the triggering strategy, one can stabilize the plant with any positive data payload transmission rate. These results are extended in \cite{linsenmayer2017delay} to a large class of triggering strategies. 
The work~\cite{ling2017bit} provides a sufficient data payload rate for second-order systems with real eigenvalues. 
While in these works the delay is assumed to be sufficiently small to achieve stabilization,  }~\cite{OurJournal1}    considers arbitrary transmission delays in the communication \magenta{network} and quantifies the information contained in the timing of the events for the stabilization of scalar plants \textit{without disturbances}.  
In ~\cite{OurJournal1} it is shown that for small delay values   stability can be achieved with any positive information transmission rate (the rate at which sensor transmits data
payload). However, as the delay increases to values larger than a
critical threshold, the timing information contained in the triggering action itself may not be enough to stabilize the plant and  the information transmission rate must be increased. The results in~\cite{OurJournal1} are valid for   vector plants when the open-loop gain matrix has only real eigenvalues.

The literature has not considered to what extent the implicit timing
information in the triggering events is useful in the presence
of plant disturbances \magenta{for the whole spectrum of possible
  bounded communication delays}. Beyond the uncertainty due to the
unknown delay in communication, disturbances add an additional degree
of uncertainty to the state estimation process.  \magenta{The required rate for
  stabilization and the viable notion for stabilization over
  communication channels critically depend  on the presence of
  disturbances~\cite{nair2004stabilizability,sahai2006necessity,matveev2009estimation,khojasteh2018stabilizing}.
  With this in mind, and in contrast to~\cite{OurJournal1} that requires
  exponential convergence guarantees, here we study
  input-to-state practical stability
  (ISpS)~\cite{jiang1994small,sharon2012input} of a linear,
  time-invariant plant subject to bounded disturbances over a communication channel with arbitrarily large but bounded~delay.}

Our  contributions  are threefold. First, for scalar real plants with
disturbances, we derive a \textit{sufficient} condition on the information
transmission rate for the whole spectrum of possible
communication delay values. Specifically, we design an encoding-decoding scheme
that, together with the proposed event-triggering strategy, rules out
Zeno behavior and ensures that there
exists a control policy which renders the plant ISpS.
We show that for small values of the delay, our event-triggering
strategy achieves 
ISpS
using only implicit timing information
and transmitting data payload at a rate arbitrarily close to zero.  On
the other hand, since larger values of the delay imply that the
information transmitted has become excessively outdated and corrupted by the
disturbance, increasingly higher communication
rates are required as the delay becomes larger.
Our second contribution pertains to the generalization of the
 sufficient  condition to complex plants with complex open-loop
gain subject to disturbances. This result sets the basis for the
generalization of event-triggered control strategies that meet the
bounds on the information transmission rate for the ISpS of vector
systems under disturbances and with any real open-loop gain matrix
(with complex eigenvalues).  The first two contributions provide
stronger results than our preliminary conference
papers~\cite{princeton_paper,cdc18paper} and contain a complete
technical treatment.  Our final  contribution is a
\textit{necessary} condition on the information transmission rate for
scalar real plants, assuming that at each triggering time the sensor
transmits the smallest possible packet size to achieve the triggering
goal for all realizations of the delay and plant disturbance. The simulation results are presented in Appendix~\ref{sec:sim}.


\section{Problem formulation}\label{sec:setup}
\begin{figure}[htb]
	\centering
    \includegraphics[scale=0.3]{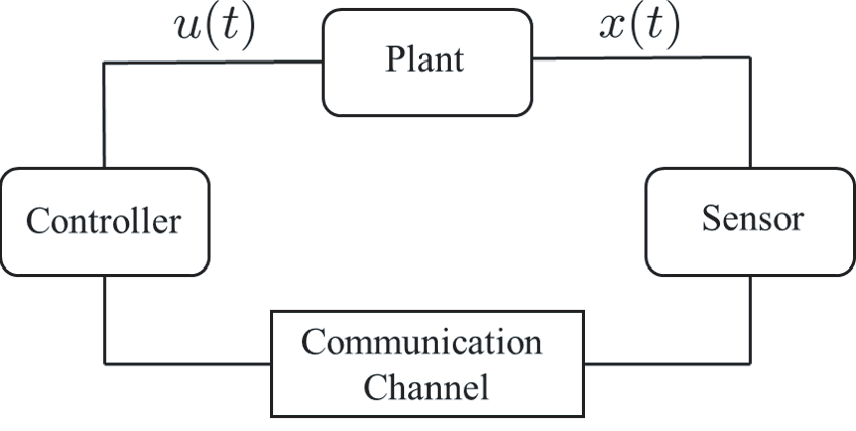}
	\caption{Networked control system model.}\label{fig:system}
	\vspace*{-2ex}
\end{figure}

We consider\footnote{
Throughout the paper, we use the following notation. $\real$, $\real_{\ge 0}$, $\complex$, and
$\integers$ represent the set of real, nonnegative real, complex, and natural numbers,
resp. We let $|.|$ and $\|.\|$ denote absolute value and complex absolute value,
resp. Let $\log$ and $\ln$ represent base $2$ and natural logarithms,
resp. For a function $f : \real \rightarrow \real^n$ and $t \in
\real$, we let $f(t^+) = \lim_{s \rightarrow t^+} f(s)$ denote the
right-hand limit of $f$ at~$t$.  In addition, $\floor{x}$
(resp. $\ceil{x}$) denotes the nearest integer less (resp. greater)
than or equal to $x$. We denote the modulo function by $\modulo(x,y)$,
representing the remainder after division of $x$ by~$y$. The function
$\text{sign}(x)$ denotes the sign of~$x$. Any $Q \in \complex$ can be
written as $Q=\RE(Q)+i\IM(Q) =\|Q\|e^{i\phi_Q}$, and for any $y \in
\real$ we have $\|e^{Qy}\|=e^{\RE(Q)y}$. 
$\Tr(A)$ denotes the trace of matrix $A$, and
$m$ denotes the Lebesgue measure. 
For a scalar continuous-time signal $w(t)$, we define $|w|_t=\sup_{s \in [0,t]} |w(s)|$.
To formulate the stability properties, for non-negative constant $d$ we define
\begin{align}
    &\mathcal{K}(d):= 
    \{f:\real_{\ge 0} \rightarrow \real_{\ge 0} \;|\; f~\mbox{continuous,}
     \\
        &~~~~~~~~~~~~~~~~
    \mbox{strictly
increasing, and}~f(0)=d\},
\\
&\mathcal{K_\infty}(d):= 
    \{f \in \mathcal{K}(d)| f~\mbox{unbounded}\},
    \\
         &\mathcal{K}_\infty^2:= 
    \{f:\real_{\ge 0} \times \real_{\ge 0} \rightarrow \real_{\ge 0} \;| \; \forall t > 0, \\
    &~~~~~~~~~~~~~ f(.,t) \in \mathcal{K}_\infty(0),~\mbox{and}~\forall r> 0~f(r,.) \in \mathcal{K}_\infty(0)\}
    \\
    &\mathcal{L}:= 
    \{f:\real_{\ge 0} \rightarrow \real_{\ge 0} \;|\; f~\mbox{continuous,}\\
    &~~~~~~~~~~~~~~~~\mbox{strictly
    decreasing, and}~\lim_{s \rightarrow \infty} f(s)=0\},
    \\
     &\mathcal{KL}:= 
    \{f:\real_{\ge 0} \times \real_{\ge 0} \rightarrow \real_{\ge 0} \;|\; f~\mbox{continuous,}\\
    &~~~~~~~~~~~~~~~~\forall t \ge 0, f(.,t) \in \mathcal{K}(0),~\mbox{and}~\forall r>0~f(r,.) \in \mathcal{L}\}.
\end{align}
} a \magenta{NCS} described by a
plant-sensor-channel-controller tuple, cf.~\figref{fig:system}.
The  plant is described by a scalar, continuous-time, linear time-invariant model,
\begin{align}\label{syscon}
   \dot{x}=Ax(t)+Bu(t)+w(t),
\end{align}
where $x(t) \in \real$ and $u(t) \in \real$ for $t \in [0,\infty)$ are the plant state and control input, respectively, and $w(t)  \in \real$ represents the plant disturbance. The latter is a Lebesgue-measurable function of time, and \textit{upper bounded} as
\begin{align}\label{noiseupp}
|w(t)|\le M,
\end{align}
where $M \in \real_{\ge 0}$. In~\eqref{syscon}, $A \in \real$ is  positive (i.e., the plant is unstable),  $B \in \real \setminus\{0\}$,
and the initial condition $x(0)$ is bounded. 
We assume the sensor measurements are exact and there is no delay in
the control action, which is executed with infinite
precision. However, measurements are transmitted from sensor to
controller over a communication channel subject to a finite data rate
and bounded unknown delay\footnote{\magenta{In general, there might
    also be a communication channel with finite capacity in the
    downlink, between the controller and the plant. However, in many
    applications such as mobile robots~\cite{siegwart2011introduction},
    the uplink, which is studied here, is the main bottleneck, as a
    strong on-board transmitter reduces the operating duration,
    restricts robot mobility, and increases cost.}}.  We denote by
$\{ t_s^k \}_{k \in \integers}$ the sequence of times when the sensor
transmits a packet of length $g(t_s^k)$ bits containing a quantized
version of the encoded state. We let $\Delta'_k = t_s^{k+1}-t_s^{k}$
be the $k^{th}$ \emph{triggering interval}. The packets are delivered
to the controller without error and entirely but with unknown upper
bounded delay. Let $\{ t_c^k \}_{k \in \integers}$ be the sequence of
times where the controller receives the packets transmitted at times
$\{ t_s^k \}_{k \in \integers}$. We assume the \emph{communication
  delays} $\Delta_k = t_c^k-t_s^k $, for all $k \in \integers$,
satisfy
\begin{align}\label{gammma}
 \Delta_k\leq \gamma ,
\end{align}
where $\gamma \in \real_{\ge 0}$. When referring to a generic
triggering or reception time, for convenience, we skip the super-script
$k$ in $t_s^k$ and~$t_c^k$, and the sub-script $k$ in $\Delta_k$
and~$\Delta'_k$.  In our model, clocks are synchronized at the sensor
and the controller. In case of using a timestamp, due to the
communication constraints, only a quantized version of it can be
encoded in the packet $g(t_s)$.

At the controller, the estimated state is represented by $\hat{x}$ and evolves during the inter-reception times as
\begin{align}\label{sysest}
  \dot{\hat{x}}(t)=A\hat{x}(t)+Bu(t), \quad t \in (t_c^k,t_c^{k+1}),
\end{align}
starting from $\hat{x}(t_c^{k+})$, which represents the state estimate of the controller with the information received up to time $t_c^k$ with initial condition $\hat{x}(0)$ (the exact way to construct $\hat{x}(t_c^{k+})$ is explained later in Section~\ref{sec:design}).
%

\begin{assume}\label{xhatassump!}
  The sensor can compute {\blue{$\{\hat{x}(t_c^{k+}) \}_{k \in \integers}$.}}
\end{assume}
\begin{remark}
\label{eee342!!!!2}
{\rm 
We show in Proposition~\ref{lemmaoffeedback} that Assumption~\ref{xhatassump!} is valid for our controller design, {\blue{provided the sensor knows 
the times the actuator performs the control action.}}
This is a common practice in TCP-based networks, where packet arrivals are
acknowledged via a communication feedback link, to ensure the robust
transmission of the packets, see
e.g.~\cite{acknowledgement_fdbk,you2010minimum,
yuksel2012random,gupta2009data}.}
In \magenta{NCS}, this corresponds to assuming an  instantaneous acknowledgment from the actuator to the sensor via the control input, known as \emph{communication through the control input}~\cite{sahai2006necessity,matveev2009estimation,tatikonda2004control}.
To obtain such causal knowledge, one can monitor the output of the actuator provided that the control input changes at each reception time.  In case the sensor has only access to the plant state, since the system disturbance is bounded~\eqref{noiseupp}, assuming that the control input is continuous during inter-reception times and jumps in the reception times such that $B|u(t_c)-u(t_c^-)|>M$, the controller can signal the reception time of the packet to the sensor via $\dot{x}(t)$. Finally, we note that any necessary condition on the information transmission rate obtained with Assumption~\ref{xhatassump!} in place remains necessary without it as well \magenta{(cf. Section~\ref{sec:estim-err-necessary})}.
  \oprocend
\end{remark}

{\blue{Under Assumption~\ref{xhatassump!}, the sensor can use~\eqref{sysest} to compute $\hat{x}(t)$ for all $t \ge 0$, provided it knows $\hat{x}(0)$. Thus, under this assumption}}, the \emph{estimation error} at the sensor~is
\begin{align}\label{eq:state-estimation-error}
z(t)=x(t)-\hat{x}(t),
\end{align}
and we rely on this error to determine when a triggering event occurs
in our controller design. We next define a modified version of
input-to-state practical stability
(ISpS)~\cite{jiang1994small,sharon2012input}, which is suitable for
the present setup.

\begin{defn}\label{cd12emkmkd}
  The plant~\eqref{syscon} is ISpS if there exist $\xi \in \mathcal{KL}$, $\psi \in \mathcal{K}_\infty(0)$, $d \in \real_{\ge 0}$, 
$\iota \in \mathcal{K}_\infty(d)$,
and $\vartheta \in \mathcal{K}_\infty^2$
 such that
\begin{align}\label{kkii34i444!222}
    |x(t)| \! \le \! \xi\left(|x(0)|,t\right) \!+\!\psi\left(|w|_t\right)\!+\!\iota(\gamma)\!+\! \vartheta(|w|_t,\gamma), \; \forall t \ge 0.
\end{align}
\end{defn}
%
%

Note that, for a fixed $\gamma$, this definition reduces to the standard notion of ISpS. Given that the initial condition, delay, and system disturbances are bounded, {\blue{ISpS implies that the state must be bounded at all times.}} 
Our objective is to ensure the dynamics~\eqref{syscon}  is ISpS
given the constraints posed by the system model of~\figref{fig:system}. Let $b_s(t)$ be the number of bits transmitted in the data payload by the sensor up to time $t$. The \emph{information transmission rate} is
\begin{align}
  \label{Tx-rate}
  R_s = \limsup_{t \rightarrow \infty}(b_s(t)/t)= \limsup_{N \rightarrow \infty} \Big(\sum_{k=1}^{N} g(t_s^k)\Big/\sum_{k=1}^{N}
   \Delta'_k\Big),~~
\end{align}
where the latter equality follows by noting that, at each triggering time $t_s^k$, the sensor transmits $g(t_s^k)$ bits.

In addition to  the data payload, the reception time of the packets carries information. Consequently,
let $b_c(t)$ be the amount of information measured in bits included in data payload and timing information 
received at the controller until time $t$.   
The \emph{information access rate} is $ R_c = \limsup_{t \rightarrow \infty} (b_c(t)/t)$.

\begin{remark}{\rm 
 We do not consider the bounded delays~\eqref{gammma} to be chosen from any specific distribution. Thus, the information that can be gained about the triggering time $t_s$ from the reception time $t_c$ may be quantified by the R\'{e}nyi 0th-order information functional $I_0$~\cite{girish2013,shingin2012disturbance}. Assuming the controller has received  $N$ packet by time $t$, we deduce $b_c(t)=\sum_{k=1}^{N} \left(g(t_s^k)+I_0(t_s^k;t_c^k)\right)$.}
  \oprocend
\end{remark}

According to the data-rate theorem~\cite{hespanha2002towards,OurJournal1}, if $R_c < A/\ln2$, the value of the state in~\eqref{syscon} becomes unbounded as $t\rightarrow \infty$,
and hence~\eqref{syscon} is not ISpS. The data-rate theorem characterizes what is needed by the controller, and does not depend on the specific feedback structure (including aspects such as information pattern at the sensor/controller, communication delays, and whether transmission times are state-dependent, as in event-triggered control, or periodic, as in time-triggered control). In our discussion below, the bound $R_c=A/\ln2$ serves as a baseline for our results on the information transmission rate $R_s$ to understand the amount of  timing information contained in event-triggered control designs in the presence of unknown communication delays.

We do not consider delays, plant disturbances, and initial condition to be chosen from any specific distribution. Therefore, our results are valid for any arbitrary delay, plant disturbances, and initial condition with finite support. In particular, our goal is to find upper and lower bounds on $R_s$, where the \textit{lower bound} is necessary at least for \textit{a realization} of the initial condition, delay, and disturbances, and the \textit{upper bound} is sufficient for \textit{all realizations} of the initial condition, delay, and disturbances. In addition, our lower bound is necessary for \textit{any} control policy $u(t)$ to render the plant~\eqref{syscon} ISpS under the class of event-triggering strategies described next.

\section{Event-triggered design}\label{sec:design}

Here we introduce the general class of event-triggered policies considered in this paper.
%
%
Consider the following class of triggers: for $J \in \real$ positive, the sensor sends a message to the controller at $t_s^{k+1}$ if
\begin{align}\label{eq:ets}
  |z(t_s^{k+1})|=J,
\end{align}
provided $t_c^k \le t_s^{k+1}$ for $k \in \mathbb{N}$ and $t_s^1 \ge 0$. A new transmission happens only after the previous packet has been received by the controller. 
Since the triggering time $t_s$ is a real number, its knowledge can reveal an unbounded amount of information to the controller. However, due to the unknown delay in the \magenta{communication network}, the controller does not have perfect knowledge of it. In fact, both the finite data rate and the delay mean that the controller may not be able to compute the exact value of~$x(t_c)$.  To address this, let $\bar{z}(t_c)$ be an estimated version of ${z}(t_c)$ reconstructed by the controller knowing $|z(t_s)|=J$, the bound~\eqref{gammma} on the delay, and the packet received through the channel. 
Using $\bar{z}(t_c)$,
the controller updates the state estimate via the \emph{jump strategy},
\begin{align}
 \hat{x}(t_c^+)=\bar{z}(t_c)+\hat{x}(t_c).
 \label{eq:jumpst}
\end{align}
Note that $|z(t_c^+)|=|x(t_c)-\hat{x}(t_c^+)|=|z(t_c)-\bar{z}(t_c)|$.
We assume the packet size $g(t_s)$ calculated at the sensor is so that
\begin{align}
|z(t_c^+)| = |z(t_c)-\bar{z}(t_c)| \le  J,
  \label{eq:jump-upp}
\end{align}
is satisfied for all $t_c \in [t_s,t_s+\gamma]$.  This property plays
a critical role in our forthcoming developments. In particular, we
will show that our controller design for the sufficient
characterization on the transmission rate is based  on identifying a
particular encoding-decoding strategy and a packet size to
ensure~\eqref{eq:jump-upp}. Likewise, our necessary characterization
is based on identifying the minimal packet sizes necessary to
ensure~\eqref{eq:jump-upp}.

The importance of~\eqref{eq:jump-upp} starts to become apparent in the following result: if this inequality holds at each reception time, the state estimation error~\eqref{eq:state-estimation-error} is bounded for all time.

\begin{lemma}\label{lem:uperboundz}
  Consider the  model with plant dynamics~\eqref{syscon}, estimator dynamics~\eqref{sysest}, triggering strategy~\eqref{eq:ets}, and jump strategy~\eqref{eq:jumpst}. 
  Assume $|z(0)|=|x(0)-\hat{x}(0)|<J$ and~\eqref{eq:jump-upp} holds at all reception times $\{t^k_c\}_{k \in \mathbb{N}}$.
  Then, for all $t\ge 0$,
  \begin{align}\label{upponz}
    |z(t)|\le J e^{A\gamma}+\frac{|w|_t}{A} \left(e^{A\gamma}-1\right).
  \end{align}
\end{lemma}
\begin{IEEEproof}
  At the reception time, $z(t_c^{k+})$ satisfies~\eqref{eq:jump-upp},
  hence using the triggering rule~\eqref{eq:ets}, we deduce $|z(t)|\le
  J$ for all $t \in (t_c^{k},t_s^{k+1}]$.  Since $J$ is smaller than
  the upper bound in~\eqref{upponz}, and $z(t_c^{(k+1)+})$
  satisfies~\eqref{eq:jump-upp}, it remains to prove~\eqref{upponz}
  for $t \in
  (t_s^{k+1},t_c^{k+1})$. From~\eqref{syscon},~\eqref{sysest},
  and~\eqref{eq:state-estimation-error}, we have
  ${\dot{z}(t)=Az(t)+w(t)}$ during inter-reception time intervals
  $(t_c^k,t_c^{k+1})$. Also, from~\eqref{eq:ets} it follows
  $(t_s^{k+1},t_c^{k+1}) \subseteq (t_c^k,t_c^{k+1})$. Thus, for all
  $t \in (t_s^{k+1},t_c^{k+1})$, we have
  \begin{align}\label{solz}
  z(t)=e^{A(t-t_s^{k+1})}z(t_s^{k+1})+\int_{t_s^{k+1}}^{t} e^{A(t-\tau)}w(\tau) d\tau.
  \end{align}  
When a triggering occurs $|z(t_s^{k+1})|=J$, hence 
the absolute value of the first addend in~\eqref{solz} is upper bounded by
$ Je^{A(t-t_s^{k+1})}$. Also, 
for
the second addend in~\eqref{solz} we have
  \begin{align}\label{calculation}
    &|\int_{t_s^{k+1}}^{t}e^{A(t-\tau)}w(\tau)d\tau|
    \\
   \nonumber
   & \quad \le |w|_t\int_{t_s^{k+1}}^{t}|e^{A(t-\tau)}|d\tau=
   \frac{|w|_t}{A} \left(e^{A(t-t_s^{k+1})}-1\right).
   \end{align}
   By~\eqref{gammma}, $t-t_s^{k+1} \!\le\! t_c^{k+1}-t_s^{k+1} \! \le \!
   \gamma$, and the result follows.
\end{IEEEproof}

{\blue{We continue by showing that, if~\eqref{eq:jump-upp} holds at each reception time $\{t_c^k\}_{k\in \mathbb{N}}$, then 
a linear controller 
renders the plant~\eqref{syscon} ISpS. We note that similar 
results exist in the literature (e.g.,~\cite{peralez2018event,heemels2012periodic,postoyan2011unifying})
and we extend them here to our event-triggering setup with quantization and unknown delays.

\begin{Proposition}\label{practicalstb}
Under the assumptions of Lemma~\ref{lem:uperboundz},
the controller~$u(t)=-K \hat{x}(t)$~renders~\eqref{syscon}~ISpS,~provided~$A-BK<0$. 
\end{Proposition}
\begin{IEEEproof}
By letting  ${u(t)=-K(x(t)-z(t))}$, we rewrite~\eqref{syscon} as $\dot{x}(t)=(A-BK)x(t)+BKz(t)+w(t)$.
Consequently,
   \begin{align}\label{!!!!!2344ngkfkef}
  |x(t)|& \le e^{(A-BK)t}|x(0)|\\\nonumber
  & 
\quad  +e^{(A-BK)t} \int_{0}^{t}e^{-(A-BK)\tau} (BK|z(\tau)|+|w(\tau)|)d\tau .
  \end{align}
  since $A-BK<0$, the first summand in~\eqref{!!!!!2344ngkfkef} is a $\mathcal{KL}$ function of $|x(0)|$ and time. Thus, it remains to prove the second summand in~\eqref{!!!!!2344ngkfkef} is upper bounded by summation of a $\mathcal{K}_\infty (0)$ function of $|w|_t$, a  $\mathcal{K}_\infty (d)$  function of $\gamma$, and a $\mathcal{K}_\infty^2$ function of $|w|_t$ and $\gamma$. The second summand in~\eqref{!!!!!2344ngkfkef} is upper bounded by $-(1-e^{(A-BK)t}) (BK |z|_t+|w|_t)/(A-BK)$.
  Since $1-e^{(A-BK)t}<1$, using Lemma~\ref{lem:uperboundz} we deduce the second summand in~\eqref{!!!!!2344ngkfkef} is upper bounded by
$\psi\left(|w|_t\right)+\iota(\gamma)+\vartheta(|w|_t,\gamma)$,
where $\psi(|w|_t)=(|w|_t/-(A-BK))$ which is a $\mathcal{K}_\infty(0)$ function of $|w|_t$, $\iota(\gamma)=((BKJe^{A\gamma})/-(A-BK))$ which is a $\mathcal{K}_\infty(d)$ function of $\gamma$ with $d=\iota(0)$,
%
%
and $\vartheta(|w|_t,\gamma)=((BK|w|_t)/-A(A-BK))(e^{A\gamma}-1)$  which is a $\mathcal{K}_\infty^2$ function of $\gamma$ and $|w|_t$.
\end{IEEEproof}
}}
Using~\eqref{noiseupp}, we deduce from Lemma~\ref{lem:uperboundz} that  $|z(t)|\le J e^{A\gamma}+\frac{M}{A} \left(e^{A\gamma}-1\right)$ for all $t\ge 0$.
Next, we rule out Zeno behavior (an infinite amount of events in a
finite time interval) for our event-triggered control design. To
do this, let $0<\rho_0<1$ be a design parameter, and assume the packet
size $g(t_s)$ is selected at the sensor to ensure a stronger version
of~\eqref{eq:jump-upp},
\begin{align}
|z(t_c^+)| = |z(t_c)-\bar{z}(t_c)| \le \rho_0 J.
  \label{eq:jump-upp-suffpart}
\end{align}
Clearly,~\eqref{eq:jump-upp-suffpart}
implies~\eqref{eq:jump-upp}. Next, we show that
given~\eqref{eq:jump-upp-suffpart}, the time between consecutive
triggers is uniformly lower bounded.

\begin{lemma}\label{le:mine}
  Consider the  model with plant dynamics~\eqref{syscon}, estimator dynamics~\eqref{sysest}, triggering strategy~\eqref{eq:ets}, and jump strategy~\eqref{eq:jumpst}. Assume $|z(0)|=|x(0)-\hat{x}(0)|<J$ and~\eqref{eq:jump-upp-suffpart} holds at all reception times $\{t^k_c\}_{k \in \mathbb{N}}$. Then for all $k \in \mathbb{N}$
$t_s^{k+1}-t_s^k \ge \ln \Big(\frac{JA+M}{\rho_0JA+M} \Big)\Big/A.$
\end{lemma}
\begin{IEEEproof}
By considering two successive triggering times $t_s^k$ and $t_s^{k+1}$ and the reception time $t_c^k$, from~\eqref{eq:ets} it follows $t_s^k \leq t_c^k \leq t_s^{k+1}$. From~\eqref{syscon},~\eqref{sysest}, and~\eqref{eq:state-estimation-error}, we have ${\dot{z}(t)=Az(t)+w(t)}$ during inter-reception time intervals $(t_c^k,t_c^{k+1})$, consequently using the definition of
the triggering time $t_s^{k+1}$~\eqref{eq:ets} it follows $|z(t_c^{k+}) e^{A(t_s^{k+1}-t_c^k)}|+|\int_{t_c^{k}}^{t_s^{k+1}} e^{A(t_s^{k+1}-\tau)}w(\tau) d\tau|\ge J$.
Using~\eqref{eq:jump-upp-suffpart} and~\eqref{calculation}, we have  $\rho_0Je^{A(t_s^{k+1}-t_c^k)}+(M/A) (e^{A(t_s^{k+1}-t_c^k)}-1)\ge J$,
which is equivalent to 
$t_s^{k+1}-t_c^k \ge \frac{1}{A} \ln (\frac{J+\frac{M}{A}}{\rho_0J+\frac{M}{A}})
$. The result follows from using~$t_s^k \le t_c^k $ in this inequality. 
\end{IEEEproof}

Given the uniform lower bound on the inter-event time in
Lemma~\ref{le:mine}, we deduce that the event-triggered control design
does not exhibit Zeno behavior. The frequency of transmission events
is captured by the \emph{triggering rate}
\begin{align}\label{trate} 
R_{tr} &= \limsup_{N\rightarrow
    \infty}\Big(N\Big/\sum_{k=1}^N \Delta'_k\Big).
\end{align}
Using Lemma~\ref{le:mine}, we deduce that the triggering
rate~\eqref{trate} is uniformly upper bounded under the
event-triggered control design, i.e., for \textit{all} initial
conditions, possible delay and plant noise values,
\begin{align}\label{uppertrig1234}
  R_{tr} \le A\Big/\ln \Big(\frac{JA+MA}{\rho_0JA+M}\Big).
\end{align} 
 \vspace*{-5ex}
\section{Sufficient and necessary conditions on the information transmission rate}\label{sec:estim-err}

Here we derive sufficient and necessary conditions on the information transmission rate~\eqref{Tx-rate} to ensure
\eqref{syscon} is ISpS.
As mentioned above, our approach is based on the characterization of the transmission rate required to ensure that~\eqref{eq:jump-upp} holds at all reception times.
%
%
Section~\ref{sec:Stability} introduces a quantization policy that, together with the event-triggered scheme, provides a complete control design to 
guarantee~\eqref{syscon} is ISpS and rules out Zeno behavior.
Section~\ref{sec:estim-err-necessary} presents 
lower bounds on the packet size and triggering rate required to 
guarantee~\eqref{syscon} is ISpS, leading to our bound on the necessary information transmission rate. We conclude the section by comparing the sufficient and necessary bounds, and discussing their gap. 
\vspace*{-3ex}
\subsection{Sufficient information transmission rate}\label{sec:Stability}

\subsubsection{Design of quantization policy}\label{sec:realn!}
The result in Proposition~\ref{practicalstb} justifies our strategy to obtain a sufficient condition on the transmission rate to guarantee~\eqref{syscon} is ISpS, which consists of finding conditions to achieve~\eqref{eq:jump-upp} for all reception times. Here we specify a quantization policy and determine the resulting estimation error as a function of the number of bits transmitted. This allows us to determine the packet size that ensures~\eqref{eq:jump-upp-suffpart} (and consequently~\eqref{eq:jump-upp}) holds, thereby leading to a complete control design which ensures \eqref{syscon} is ISpS and rules out Zeno behavior. In turn, this also yields a sufficient condition on the information transmission rate. In our 
{\blue{particular}}
design the controller estimates $z(t_c)$ as 
\begin{align}\label{ctrlapp1}
   \bar{z}(t_c)=\text{sign}(z(t_s)) J e^{A(t_c-q(t_s))},
\end{align}
where $q(t_s)$ is an estimation of the triggering time $t_s$ constructed at the controller as described next.
According to~\eqref{eq:ets}, at every triggering event, the sensor encodes $t_s$ and transmits a packet $p(t_s)$. The packet  $p(t_s)$ consists of $g(t_s)$ bits of information and is generated according to the following quantization policy.
The first bit $p(t_s)[1]$ denotes the sign of $z(t_s)$.
%
\begin{figure}[thb]
  \centering
  \includegraphics[scale=0.47]{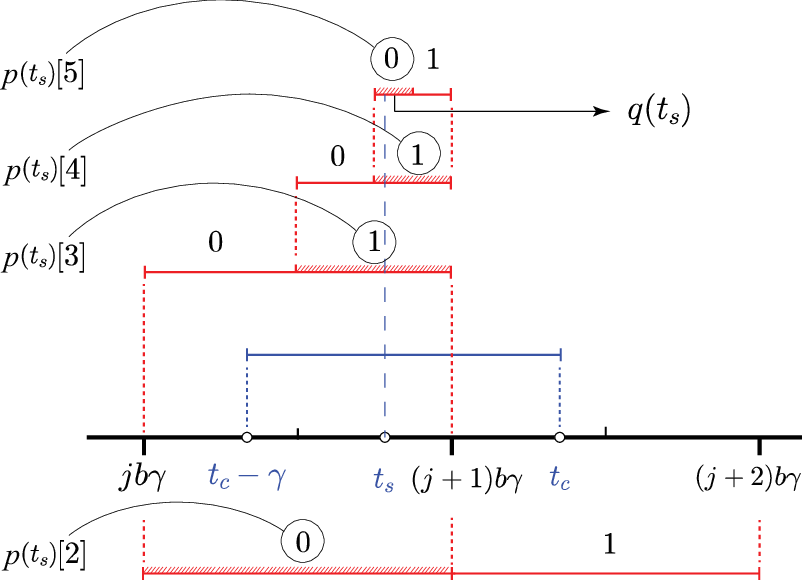} 
  \caption{The encoding-decoding algorithms in the proposed event-triggered control scheme. 
    Here,  $g(t_s)=5$ and $j$ is an even natural number. The packet
    $p(t_s)$ of length 5 can be generated and sent to the controller ($p(t_s)[1]$ encodes the sign of $z(t_s)$). After reception and
    decoding, the controller chooses the center of the smallest
    sub-interval as its estimation of $t_s$, denoted by $q(t_s)$.}
  \label{encoder}
	\vspace*{-2ex}
\end{figure}
As shown in~\figref{encoder}, the reception time $t_c$ provides information to the controller that $t_s$ could fall anywhere between $t_c-\gamma$ and $t_c$.
Let $b>1$. To determine the time interval of the triggering event, we break the positive time line into intervals of length $b\gamma$, cf. Appendix~\ref{sec:design-parameters}.
Consequently, $t_s$ falls into $[jb\gamma,(j+1)b\gamma]$ or $[(j+1)b\gamma,(j+2)b\gamma]$, with $j$ a natural number. We use the second bit of the packet to determine the correct interval of $t_s$. This bit is zero if the nearest integer less than or equal to the beginning number of the interval is an even number and is 1 otherwise. Mathematically, $p(t_s)[2]= \modulo \big( \floor{ \frac{t_s}{b\gamma} }, 2 \big)$.
For the remaining bits of the packet, the encoder breaks the interval containing $t_s$ into $2^{g(t_s)-2}$ equal sub-intervals. Once the packet is complete, it is transmitted to the controller, where it is decoded and the center point of the smallest sub-interval is selected as the best estimate of~$t_s$. Thus,
\begin{align}
	\label{Quntrulesuff1} 
    |t_s-q(t_s)| \leq b\gamma/2^{g(t_s)-1}.
\end{align}

{\blue{Pseudo-code descriptions of the above encoding and decoding algorithms are provided in Appendix~\ref{sec:pccode}.
\begin{remark}
\label{ramrk3445}
{\rm When the delay is sufficiently small, the timing information is substantial and the uncertainty about the value of the state at the controller is small. In this case,  there is no need to resort to data payloads in the packet, as the plant can be stabilized using only \textit{timing information} about the triggering events, as demonstrated in~\cite{anantharam1996bits,khojasteh2018stabilizing}. These works simply have the sensor transmits a fixed symbol from a unitary alphabet, reducing the communication channel to a \textit{telephone signaling channel} capable of stabilizing the system.} \oprocend
\end{remark}
}}
We have employed this quantization policy in our previous
work~\cite{OurJournal1} and analyzed its behavior in the case with no
disturbances. Next, we extend our analysis to scenarios with both
unknown delays and plant disturbances. As discussed in
Remark~\ref{eee342!!!!2}, we start by showing that under the proposed
encoding-decoding scheme, provided the sensor knows $\hat{x}(0)$ and
has causal knowledge of the delay (i.e., the controller acknowledges
the packet reception times), then Assumption~\ref{xhatassump!}
holds. The proof of the next result is in~Appendix~\ref{sec:proofs}.

\begin{Proposition}\label{lemmaoffeedback}
Under the assumptions of Lemma~\ref{le:mine},
using the estimation~\eqref{ctrlapp1} and the quantization policy described in~\figref{encoder}, {\blue{if the sensor 
has causal knowledge of delays, then it can calculate $\{\hat{x}(t_c^{k+}) \}_{k \in \integers}$.}}
\end{Proposition}

\subsubsection{Sufficient packet size}\label{secNecdead!}
Our next result bounds the difference  ${|t_s-q(t_s)|}$ between the triggering time and its quantized version so that~\eqref{eq:jump-upp-suffpart} holds at all reception times.

\begin{lemma}\label{lemmasu}
  Consider the  model with plant dynamics~\eqref{syscon}, estimator dynamics~\eqref{sysest}, triggering strategy~\eqref{eq:ets}, and jump strategy~\eqref{eq:jumpst}. 
  Assume $|z(0)|=|x(0)-\hat{x}(0)|<J$.
  Using the estimation~\eqref{ctrlapp1} and the quantization policy described in~\figref{encoder}, if $|t_s-q(t_s)| \le \frac{1}{A} \ln(1+\frac{\rho_0-\frac{M}{JA}(e^{A\gamma}-1)}{e^{A\gamma}})$,
  then~\eqref{eq:jump-upp-suffpart} holds for all reception times $\{t_c^k\}_{k \in \mathbb{N}}$ if $J> \frac{M}{A\rho_0}(e^{A\gamma}-1)$.
\end{lemma}
\begin{IEEEproof}
Using~\eqref{solz},~\eqref{ctrlapp1}, and the triangular inequality, we deduce $|z(t_c)-\bar{z}(t_c)|\le Je^{A(t_c-t_s)}|(1-e^{A(t_s-q(t_s))})|+|\int_{t_s}^{t_c}e^{A(t_c-\tau)}w(\tau)d\tau|$.
By applying the bounds~\eqref{gammma},~\eqref{noiseupp}, and~\eqref{calculation} on first and second addend respectively it follows $|z(t_c)-\bar{z}(t_c)|\le
|Je^{A\gamma}(1-e^{A(t_s-q(t_s))})|+(M/A) \left(e^{A\gamma}-1\right)$.
Therefore, ensuring~\eqref{eq:jump-upp-suffpart} reduces to
\begin{align}\label{eqmidd}
|1-e^{A(t_s-q(t_s))}|\le \eta,
\end{align}
where 
$\eta= e^{-A \gamma} (\rho_0-\frac{M}{AJ}(e^{A\gamma}-1))$. Since $J> \frac{M}{A\rho_0}(e^{A\gamma}-1)$, we have $0\le\eta<1$. 
Consequently, using~\eqref{eqmidd}, we deduce $\ln (1-\eta)/A\le t_s-q(t_s)\le \ln (\eta+1)/A$.
It follows that to satisfy~\eqref{eq:jump-upp-suffpart} for all delay values, requiring
$|t_s-q(t_s)| \le \min\{|\ln (1-\eta)|/A,\ln (1+\eta)/A\}$
suffices.
\end{IEEEproof}

The next result provides a lower bound on the packet size so that~\eqref{eq:jump-upp-suffpart} is ensured at all reception times.

\begin{theorem}\label{thm:suf-cond-ET} 
 Consider the  model with plant dynamics~\eqref{syscon}, estimator dynamics~\eqref{sysest}, triggering strategy~\eqref{eq:ets}, and jump strategy~\eqref{eq:jumpst}. 
  Assume $|z(0)|=|x(0)-\hat{x}(0)|<J$. Then
 there exists a quantization policy that achieves~\eqref{eq:jump-upp-suffpart}
for all reception times $\{t_c^k\}_{k \in \mathbb{N}}$ 
 with any  packet size
  \begin{align}\label{Sufi!!221}
  g(t_s^k) \!\ge \! \max\Big\{0,1+\log \frac{Ab\gamma}{\ln(1+\frac{\rho_0-(M/(JA))(e^{A\gamma}-1)}{e^{A\gamma}})}\Big\}~~~
  \end{align}
  where $b>1$ and  $J>\frac{M}{A\rho_0}(e^{A\gamma}-1)$.
\end{theorem}

The proof is a direct consequence of~\eqref{Quntrulesuff1} and  Lemma~\ref{lemmasu}. The combination of the upper bound~\eqref{uppertrig1234} obtained for the triggering rate and Theorem~\ref{thm:suf-cond-ET} yields a sufficient bound on the information transmission rate. To sum it up, we conclude that there exists an information transmission rate
  \begin{align}\label{inftranrate}
& R_s \le \\\nonumber
 &\frac{A}{\ln (\frac{JA+M}{\rho_0JA+M})} \max\left\{0,1+\log \frac{Ab\gamma}{\ln(1+\frac{\rho_0-(M/(JA))(e^{A\gamma}-1)}{e^{A\gamma}})}\right\},
  \end{align}
that is sufficient to ensure~\eqref{eq:jump-upp-suffpart} and, as a consequence~\eqref{eq:jump-upp},
for all reception times $\{t_c^k\}_{k \in \mathbb{N}}$. Therefore, from Proposition~\ref{practicalstb}, the bound~\eqref{inftranrate} is sufficient to ensure
the plant~\eqref{syscon} is ISpS.
%
%

{\blue{
\begin{remark}
\label{screm334}
{\rm 
The lower bound given on the packet size in~\eqref{Sufi!!221} might not be a natural number or might even be zero. We use it to properly bound in~\eqref{inftranrate} the information transmission rate $R_s$, which is a non-negative real number. For sufficiently small $\gamma$, if $g(t_s)=0$ is sufficient, the plant can be stabilized using only timing information and there is no need to put any data payload in the packet, cf. Remark~\ref{ramrk3445}.
If we do not use fixed symbols as in telephone signaling channels~\cite{khojasteh2018stabilizing}, in practice, the packet size should be a natural number. 
Therefore, we employ
\begin{align}\label{efkfek!!!45hgree}
\textstyle
   g(t_s) = \max\left\{1,\left\lceil 1+\log
       \frac{Ab\gamma}{\ln(1+\frac{\rho_0-(M/(JA))(e^{A\gamma}-1)}{e^{A\gamma}})}\right\rceil\right\},~~~
\end{align}
which is sufficient for stabilization
(and is the one used in our
 simulations of Appendix~\ref{sec:sim}).
} \oprocend
\end{remark}
}}

\subsection{Necessary information transmission rate}\label{sec:estim-err-necessary}

Here, we present a necessary condition on the information transmission rate required by \textit{any} control policy to render  plant~\eqref{syscon} ISpS under the class of event-triggering strategies described in Section~\ref{sec:design}. 
In Section~\ref{sec:Stability}, to derive a sufficient bound that guarantees~\eqref{syscon} is ISpS,
our focus has been on identifying \emph{a} quantization policy that could handle \emph{any} realization of initial condition, delay, and disturbance. Instead, the treatment here switches gears to focus on \emph{any} quantization policy, for which we identify at least \emph{a} realization of initial condition, delay, and disturbance  that requires the necessary bound on the information transmission rate.

Our strategy to provide a necessary condition for~\eqref{syscon} to be ISpS is based on the following observation. Note that, if the property~\eqref{eq:jump-upp} was not satisfied at an arbitrary reception time $t_c^k$ (i.e., $z(t_c^k)>J$), and in addition either $w(t)>0$ or $w(t)<0$ for all $t\ge t_c^k$, then $t_c^k$ would be the last triggering time as~\eqref{eq:ets} would never be satisfied again. Then, after $t_c^k$, the controller would need to estimate the inherently unstable plant in open loop. This would mean that there exists a realization of the initial condition, system
disturbances, and delay for which the absolute value of the state
estimation error grows exponentially with time. Thus, for any given
control policy, there would exist a realization for which the absolute
value of the state tends to infinity with time, and \eqref{syscon} is
not ISpS.

As a consequence of this observation, our strategy to
provide a necessary condition consists of identifying a necessary condition on the information transmission
rate~$R_s$ to have~\eqref{eq:jump-upp} at all reception times
$\{t_c^k\}_{k \in \mathbb{N}}$. In turn, we do this by finding lower
bounds on the packet size $g(t_s)$ and the triggering
rate~$R_{tr}$. We do this in two steps: first, we find a lower bound
on the number of bits transmitted at each triggering event which holds
irrespective of the triggering rate. Then, we find a lower bound on
the triggering rate, and the combination leads us to the necessary
condition on~$R_s$.

\subsubsection{Necessary packet size}\label{secNe!cee2!}
We rely on~\eqref{solz} to define the uncertainty set of the sensor about the estimation error at the controller $z(t_c)$ given $t_s$ as follows
\begin{align*}
  \Omega(z(t_c) | t_s) &=\{y: y=\pm Je^{A(t_r-{t}_s)}+
 \int_{t_s}^{t_r} e^{A(t_r-\tau)}w(\tau) d\tau, 
  \\\nonumber
  &\quad t_r \in
  [t_s,t_s+\gamma],~|w(\tau)|\le M~~ \text{for} ~\tau \in [t_{s},t_{r}] \}.
\end{align*}
Additionally, we define the uncertainty of
the controller about $z(t_c)$ given $t_c$, as follows
\begin{align*}
  \Omega(z(t_c) |t_c) &= \{y: y=\pm Je^{A(t_c-t_r)}+
  \int_{t_r}^{t_c} e^{A(t_c-\tau)}w(\tau) d\tau, 
  \\\nonumber
& \quad t_r \in [t_c-\gamma,t_c],~|w(\tau)|\le M~~ \text{for} ~\tau \in [t_r,t_{c}]\}.
\end{align*}
We next show the relationship between these uncertainty sets.

\begin{lemma}\label{inclusionsets}
  Consider the model described in Section~\ref{sec:setup}, with plant
  dynamics~\eqref{syscon}, estimator dynamics~\eqref{sysest},
  triggering
  strategy~\eqref{eq:ets}, 
  and jump strategy~\eqref{eq:jumpst}. Moreover, assume $M \le
  AJ$. Then $\Omega(z(t_c) |t_s)=\Omega(z(t_c)|t_c)$ and
  $m\left(\Omega(z(t_c) |t_c)\right)
  = 2 (M/A+J)(e^{A\gamma}-1)$.
\end{lemma}
\begin{IEEEproof}
Due to symmetry, one can show that $\Omega(z(t_c) |t_s)$ is the same as $\Omega(z(t_c) |t_c)$.
We characterize the set $\Omega(z(t_c) |t_s)$ as follows. We reason for the case when $z(t_s)=J$ (the argument for  $z(t_s)=-J$ is analogous). Clearly, $z(t_c)$ takes its largest value when $t_c=t_s+\gamma$ and $w(\tau)=M$ for $\tau \in [t_s,t_{c}]$, which is equal to $z(t_c)=Je^{A\gamma}+(M/A)(e^{A\gamma}-1)$.
Finding the smallest value of $z(t_c)$ is more
challenging.  When $t_c=t_s$,
\begin{align}\label{favorminztc}
z(t_c)=J.
\end{align}
 By setting $w(\tau)=-M$ for $\tau \in [t_s,t_{c}]$ and $t_c=t_s+\Delta$,
\begin{align}\label{funcdelta}
z(t_c)=Je^{A\Delta}-(M/A)(e^{A\Delta}-1).
\end{align}
Taking the derivative of~\eqref{funcdelta} with respect to $\Delta$ results in
\begin{align}\label{derivi1}
dz(t_c)/d\Delta=AJe^{A\Delta}-Me^{A\Delta}=e^{A\Delta}(AJ-M).
\end{align}
If $M \le AJ$ and the derivative in~\eqref{derivi1} is non-negative, $z(t_c)$ in~\eqref{funcdelta} would be a non-decreasing function of $\Delta$. Hence, the smallest value of $z(t_c)$ in~\eqref{funcdelta} occurs for $\Delta=0$ which is equal to the value of $z(t_c)$ in~\eqref{favorminztc}. 
Hence, $ \Omega(z(t_c) |t_s)=[J,Je^{A\gamma}+(M/A)(e^{A\gamma}-1)]$, and the result follows.
\end{IEEEproof} 

Lemma~\ref{inclusionsets} allows us to find a lower bound on the packet size~$g(t_s)$, which is valid irrespective of the triggering rate. 

\begin{lemma}\label{thm:necc-cond-ET-g}
   Under the assumptions of Lemma~\ref{inclusionsets}, if
  \eqref{eq:jump-upp} holds for all reception times $\{t_c^k\}_{k \in \mathbb{N}}$, then the
  packet size at every triggering event must satisfy 
  \begin{align}\label{g'eq}
    g(t_s^k) \ge \max \left\{0,\log \left(\left(M/(AJ)+1\right)\left(e^{A \gamma}-1\right)\right)\right\}.
  \end{align}
\end{lemma}
\begin{IEEEproof}
  To ensure \eqref{eq:jump-upp} for all reception times, we calculate
  a lower bound on the number of bits to be transmitted to ensure the
  sensor uncertainty set $\Omega(z(t_c) | t_s)$ is covered by
  quantization cells of measure $2 J$. Therefore, we have $ g(t_s) \ge
  \max \left\{0,\log \left(m(\Omega(z(t_c) | t_s))/m(\mathcal{B}(
      J))\right)\right\}$,
  where $\mathcal{B}(J)$ is a ball centered at $0$ of radius~$J$, and
  we have incorporated the fact that the packet size $g(t_s)$ must be
  non-negative. From Lemma~\ref{inclusionsets}, $ \log
  \frac{m(\Omega(z(t_c) | t_s))}{m(\mathcal{B}(J))}\ge \log \frac{
    (M/A+J)(e^{A\gamma}-1)}{ J}$.
\end{IEEEproof}

\subsubsection{Lower bound on the triggering rate}\label{secNe!ceewidn!}
Our next step is to
determine a lower bound on the triggering rate.

\begin{lemma}\label{lowetriggerrate1}
  Under the assumptions of Lemma~\ref{inclusionsets},  for all the quantization policies which ensure~\eqref{eq:jump-upp} at all reception times $\{t_c^k\}_{k \in \mathbb{N}}$, if  there exists a delay realization $\{\Delta_k \le \alpha \}_{k \in \mathbb{N}}$, a disturbance realization, and an initial condition such that 
     \begin{align}\label{eq:deltaz!!2}
     \textstyle
    |z(t_c^{k+})|=|z(t_c^k)-\bar{z}(t_c^k)|\geq \Upsilon,
  \end{align} 
  for all $k \in \mathbb{N}$,
  {\blue{then 
  \begin{align}\label{Kower!34}
    R_{tr} \ge A\left(\ln\left( e^{A\alpha}(JA+M)\big/(\Upsilon A +M)\right)\right)^{-1},
  \end{align}
  for said delay realization, disturbance realization, and initial condition.}}
\end{lemma}
\begin{IEEEproof}
  Using the definition of the triggering time~\eqref{eq:ets},~\eqref{eq:deltaz!!2}, $t_c^k=t_s^k+\Delta_k$, and~\eqref{solz}, we have $\Upsilon e^{A(t_s^{k+1}-t_s^{k}-
      \Delta_k)}+(M/A)\big(e^{A(t_s^{k+1}-t_s^{k}-
      \Delta_k)}-1\big) \le J$,
  which is equivalent to
  \begin{align}\label{M334}
    e^{A(t_s^{k+1}-t_s^{k})}\le e^{A\Delta_k} (JA+M)\big/(\Upsilon A +M).
  \end{align}
  By hypothesis,~\eqref{eq:deltaz!!2} occurs for all $k \in \mathbb{N}$ when ${\Delta_k \le \alpha}$. Hence, by~\eqref{M334}, we upper bound the triggering intervals as
  \begin{align}~\label{interdelay2}
   \Delta'_k \!=\! t_s^{k+1}-t_s^{k} \!\le \! A^{-1} \ln \big( e^{A\alpha}(JA+M)\big/(\Upsilon A+M)\big).~~\,\,
  \end{align}
   The result follows by 
   substituting~\eqref{interdelay2} into~\eqref{trate}.
\end{IEEEproof}

If we do not  limit the collection of permissible quantization policies, a packet may carry an unbounded amount of information, which can bring the state estimation error  arbitrarily close to zero at all reception times and for all delay and disturbance values. This would give rise to a conservative lower bound on the transmission rate. Specifically, using $\Delta_k \le \gamma$, cf.~\eqref{gammma}, putting $\Upsilon=0$, and combining~\eqref{Kower!34} and~\eqref{g'eq}  we deduce there exists a delay realization, disturbance realization, and initial condition such that
\begin{align}\label{necessT!!34}
    R_s \ge A \frac{\max \left\{0,\log \left(\left(\frac{M}{AJ}+1\right)\left(e^{A \gamma}-1\right)\right)\right\}}{\ln\left( e^{A\gamma}\frac{JA+M}{M}\right)},
\end{align}
is necessary for all quantization policies.
To find a tighter necessary condition, we instead limit the collection
of permissible quantization policies.  Since
ensuring~\eqref{eq:jump-upp} at each reception time is equivalent to
dividing the uncertainty set at the controller $\Omega(z(t_c) |t_c)$
by quantization cells of measure at most $2J$, our approach is to
restrict the class of quantization policies to those that use the
minimum possible number of bits to ensure~\eqref{eq:jump-upp}.

\begin{assume}\label{Defnition11}
  We assume at each triggering time the sensor transmits the smallest
  possible packet size  to ensure~\eqref{eq:jump-upp} at
  each reception time for all initial conditions and all possible
  realizations of the delay and plant disturbance.  Moreover, to
  simplify our analysis in the encoding-decoding scheme, we choose the
  center of each quantization cell as $\bar{z}(t_c)$.
\end{assume}

Based on this assumption, the sensor brings the uncertainty about
$z(t_c)$ at the controller down to a quantization cell of measure at
most $2J$, using the smallest possible packet size. The next result,
whose proof is in~Appendix~\ref{sec:proofs}, shows that, for this class of
quantization policies, there exists a delay realization such that the
sensor can only shrink the estimation error for the controller to at
most half of~$J$ dictated by~\eqref{eq:jump-upp}.

\begin{lemma}\label{NecesaaQuanClas}
Let
$ \beta = \ln
   \left(1+2 AJ/ (AJ+M)\right) \big/A\le \gamma .$
  Under the assumptions of Lemma~\ref{inclusionsets}, for all the quantization policies ensuring~\eqref{eq:jump-upp} at all reception times $\{t_c^k\}_{k \in \mathbb{N}}$   with Assumption~\ref{Defnition11} in place, there exists
  a delay realization $\{ \Delta_k \le \beta
  \}_{k \in \mathbb{N}}$, initial condition, and plant disturbance such that
  \begin{align}\label{eq:deltaz}
    |z(t_c^{k+})|=|z(t_c^k)-\bar{z}(t_c^k)|\geq J/2.
  \end{align}
\end{lemma}

Combining Lemmas~\ref{lowetriggerrate1} and~\ref{NecesaaQuanClas},
we deduce there exists a delay realization, disturbance realization, and initial condition so that
\begin{align}\label{secondlowerboundontr!!122}
    R_{tr} \ge A \left(\ln\left(\left(1+\frac{2AJ}{AJ+M}\right)\frac{JA+M}{0.5 JA +M}\right)\right)^{-1}
\end{align}
is valid for all quantization policies that use the minimum required packet size according to Assumption~\ref{Defnition11}. 

Combining the bounds on the packet size
(cf. Lemma~\ref{thm:necc-cond-ET-g}) and on the triggering rate
(cf.~\eqref{secondlowerboundontr!!122}), we obtain the following.

\begin{figure}[tbh]
	\centering
 \includegraphics[height=4cm,width=6cm]{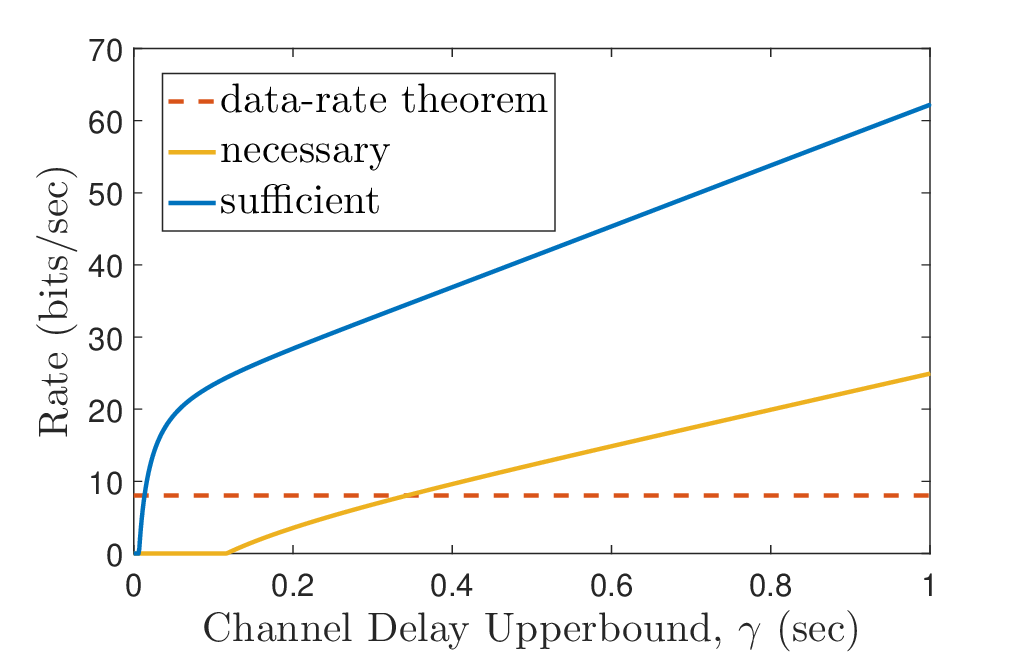}
	\caption{Illustration of the sufficient~\eqref{inftranrate} and necessary~\eqref{necessT} transmission rates as functions of the delay upper bound $\gamma$. Here, $A=5.5651$, $\rho_0=0.1$, $b=1.0001$, $M=0.4$, and $J=\frac{M}{A\rho_0}(e^{A\gamma}-1)+0.1$. The rate dictated by the data-rate theorem is $R_c \ge A/\ln2=8.02874$.
	}
    \label{fig:comparision1} 
	\vspace*{-3.5ex}
\end{figure}

\begin{theorem}\label{THM:NECESAAPP}
  Under the assumptions of Lemma~\ref{inclusionsets},
 for all the quantization policies which
   ensure
  \eqref{eq:jump-upp} at all reception times $\{t_c^k\}_{k \in \mathbb{N}}$   with Assumption~\ref{Defnition11} in place, there exists a delay realization 
  $\{ \Delta_k \le \beta
  \}_{k \in \mathbb{N}}$, a disturbance realization, and an initial condition such that 
  \begin{align}\label{necessT}
    R_s \ge A \frac{\max \left\{0,\log \left(\left(M/(AJ)+1\right)\left(e^{A \gamma}-1\right)\right)\right\}}{  \ln\left(\left(1+\frac{2 AJ }{AJ+M}\right)\frac{JA+M}{0.5 JA +M}\right)}.
  \end{align}
\end{theorem}
The bound~\eqref{necessT} is tighter than the bound
in~\eqref{necessT!!34}. \figref{fig:comparision1} compares our
bounds on the sufficient~\eqref{inftranrate} and
necessary~\eqref{necessT} information transmission rates
for~\eqref{syscon} to be ISpS. We attribute the gap between them to
the fact that, while the necessary condition employs quantization
policies with the minimum possible packet size according to
Assumption~\ref{Defnition11}, the encoding-decoding scheme in the
sufficient design does not generally satisfy this assumption.  The
fact that we bound the triggering rate and the packet size
independently in our analysis might further contribute to the gap.  The key point that is evident from \figref{fig:comparision1}, is that for sufficiently small
delay values the timing information is substantial, and the plant can
be ISpS in the presence of bounded disturbances when the sensor
transmits data payload at a smaller rate than the one prescribed by
the data-rate theorem.  As the delay increases, the timing information
becomes less useful.
Since the state estimation error is smaller than the triggering
threshold at each reception time in our design, for larger values of
delay, $R_s$ exceeds the access rate prescribed by the data-rate
theorem.


\section{Extension to complex linear systems}\label{sec:HigherDimensions}

In this section, we generalize our treatment to complex linear plants with disturbances. The results presented here can be readily applied to multivariate linear plants with disturbance and  diagonalizable open loop-gain matrix (possibly, with complex eigenvalues). This corresponds to handling the $n$-dimensional real plant as $n$ scalar (and possibly complex) plants, and derive a sufficient condition for them. We consider a plant, sensor, communication channel and controller described by the  continuous linear time-invariant system
\begin{align}
  \label{sysconHD}
  \dot{x}=Ax(t)+Bu(t)+w(t),
\end{align}
where $x(t)$ and $u(t)$ belong to $\complex$ for $t \in [0,\infty)$. Here $w(t)  \in \complex$ represents a plant disturbance, which is upper bounded as $\|w(t)\|\le M$, with $M \in \real_{\ge 0}$. Also, $A \in \mathbb{C}$ with $\RE(A) \ge 0$ (since we are only interested in unstable plants) and $B \in \complex$ is nonzero.
The model for the communication channel is the same as in
Section~\ref{sec:setup}. To establish a baseline for comparison of the
bounds on the information transmission rate, we start by stating a
generalization of the classical data-rate theorem for the complex
plant~\eqref{sysconHD}.  The proof is in~Appendix~\ref{sec:proofs}.

\begin{theorem}\label{thm:necc-access-rate-HD}
  Consider the  model with plant dynamics~\eqref{sysconHD}. If $\|x(t)\|$ remains bounded as $t \rightarrow \infty$, then $R_c \ge 2\RE(A)/\ln 2$.
\end{theorem}

\subsection{Event-triggered control for complex linear systems}\label{sec:event-trigger-complex}
\begin{figure}[thb]
	\centering
 \includegraphics[scale=0.21]{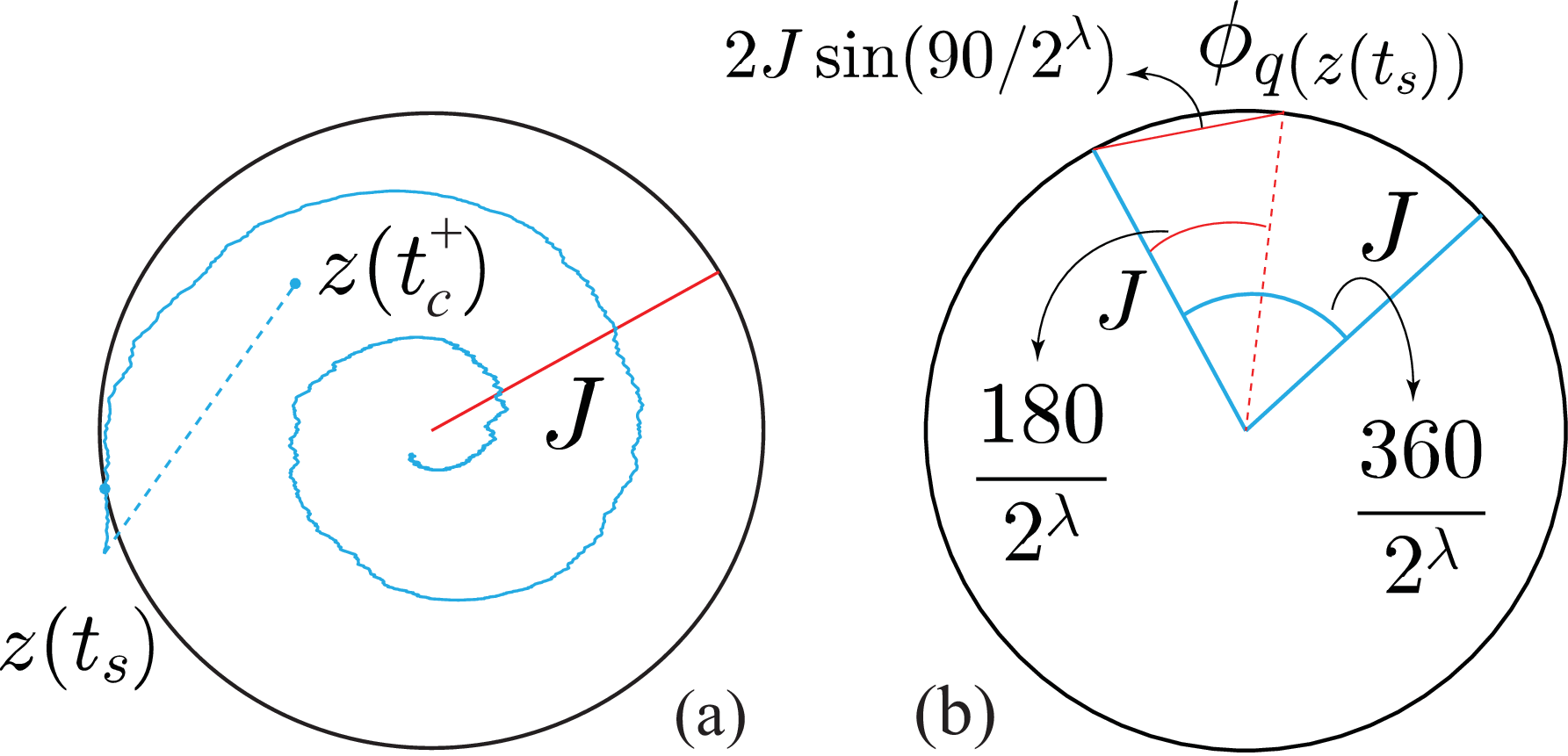}
	\caption{(a)  Evolution of the state estimation error (blue curve)  before and after an event. The trajectory starts 
    with an initial state inside a circle of radius $J$, and continues spiraling (due to the imaginary part of $A$) until it hits the  threshold~$J$. Then it jumps back inside the circle after the update according to~\eqref{zbar-tc-estimate} and jump strategy~\eqref{eq:jumpst}.
    During inter-reception time intervals, ${\dot{z}(t)=Az(t)+w(t)}$, and the observed overshoot beyond the circle is due to the delay in the communication channel. Here, $A=0.3+2i$, $B=0.2$, $u(t)=-8\hat{x}(t)$, $M=0.2$, $\gamma=0.05$ sec, $\rho_0=0.9$ and $J=0.0173$.
    (b) Estimation of the phase angle after event and transmission of $\lambda$ bits.}
    \label{fig:triggering-cricle} 
	\vspace*{-1.5ex}
\end{figure}

The state estimate $\hat{x}$ evolves according to the dynamics~\eqref{sysest} along the inter-reception time intervals starting from $\hat{x}(t_c^{k+})$ with initial condition $\hat{x}(0)$. 
We use the \emph{state estimation error} defined as~\eqref{eq:state-estimation-error}
with initial condition $z(0)=x(0)-\hat{x}(0)$.
A triggering event happens  at $t_s^{k+1}$ if
\begin{align}\label{eq:etsHHD}
 \|z(t_{s}^{k+1})\|=J,
\end{align}
provided $t_c^k \le t_s^{k+1}$ for $k \in \mathbb{N}$ and $t_s^1 \ge 0$, and the triggering radius  $J \in \real$ is positive. 
At each triggering time, the packet $p(t_s)$ of size $g(t_s)$ is transmitted from the sensor to the controller.  The packet $p(t_s)$ consists of a quantized version of the phase of $z(t_s)$, denoted $\phi_{q(z(t_s))}$, and a quantized version of the triggering time $t_s$. By~\eqref{eq:etsHHD}, we have $z(t_s) = Je^{i\phi_{z(t_s)}}$.
We construct a quantized version, denoted $q(z(t_s))$, of $z(t_s)$ at the controller as $q\left(z(t_s)\right)=Je^{i\phi_{q\left(z(t_s)\right)}}$.
%
Additionally, using the bound~\eqref{gammma} and the packet at the controller, the quantized version of $t_s$ is reconstructed and denoted by $q(t_s)$. Hence, at the controller, $z(t_c)$ is estimated as follows
\begin{align}\label{zbar-tc-estimate}
   \bar{z}(t_c)= e^{A\left(t_c-q(t_s)\right)}q\left(z(t_s)\right).
\end{align}
We use the jump strategy~\eqref{eq:jumpst} to update the value of $\hat{x}(t_c^+)$. Hence, $\|z(t_c^+)\|=\|z(t_c)-\bar{z}(t_c)\|$ holds.
At the sensor, the packet size $g(t_s)$ is chosen to be large enough such that 
\begin{align}
\|z(t_c^+)\| = \|z(t_c)-\bar{z}(t_c)\| \le \rho_0 J, 
  \label{eq:jump-upp-HD}
\end{align}
(where $0<\rho_0<1$ is a design parameter) is satisfied for all ${t_c
  \in [t_s,t_s+\gamma]}$. \figref{fig:triggering-cricle}(a) shows a
typical realization of $z(t)$ under the proposed event-triggered
strategy before and after one event. The notion of ISpS
is the same as in Definition~\ref{cd12emkmkd} by replacing absolute value with complex absolute value.
\vspace*{-0.4ex}

\begin{remark}\label{akhar12} {\rm 
  Similarly to Proposition~\ref{practicalstb}, one can show that  if~\eqref{eq:jump-upp-HD} occurs at all reception times and $(A,B)$ is a stabilizable pair, then under the control rule $u(t)=-K \hat{x}(t)$, the plant~\eqref{sysconHD} is ISpS,
  provided the real part of  $A-BK$ is negative. As a consequence of this observation, our analysis focuses on ensuring~\eqref{eq:jump-upp-HD} at each reception time. The lower bound on the inter-event time of   Lemma~\ref{le:mine}  and the upper bound on the triggering rate~\eqref{uppertrig1234} also holds replacing $A$ by $\RE (A)$.
  }       
  \oprocend
\end{remark}

\subsection{Sufficient information transmission rate}
We design a quantization policy that, using the event-triggered controller of Section~\ref{sec:event-trigger-complex}, 
ensures the plant~\eqref{sysconHD} is ISpS. We rely on this design to establish a sufficient bound on the information transmission rate.

\subsubsection{Design of quantization policy}\label{sec:evencomplexdes}
We devote the first $\lambda$ bits of the packet $p(t_s)$ for quantizing the phase of $z(t_s)$. The proposed encoding algorithm uniformly quantizes the circle into $2^{\lambda}$ pieces of $2\pi/2^{\lambda}$ radians. After reception, the decoder finds the correct phase quantization cell and selects its center point as $\phi_{q\left(z(t_s)\right)}$. By letting $\omega=\phi_{z(t_s)}-\phi_{q\left(z(t_s)\right)}$,
%
%
as depicted in \figref{fig:triggering-cricle}(b), geometrically we deduce
$|\omega|\le \pi/2^{\lambda}$.
Furthermore, we use the encoding scheme proposed in~\figref{encoder} to append a quantized version of triggering time $t_s$ of length $g(t_s)-\lambda$ to the packet $p(t_s)$.
Hence, 
$p(t_s)[\lambda+1]= \modulo \big( \floor{ \frac{t_s}{b\gamma} }, 2 \big)$.
For the remaining bits of the packet, the encoder breaks the interval containing $t_s$ into $2^{g(t_s)-\lambda-1}$ equal sub-intervals. Once the packet is complete, it is transmitted to the controller, where it is decoded and the center point of the smallest sub-interval is selected as the best estimate of $t_s$. Therefore,
\begin{align}
	\label{Quntrulesuff1C} 
    |t_s-q(t_s)| \leq b\gamma/2^{g(t_s)-\lambda}.
\end{align}
 Given $t_s^{k+1}$, one can identify $q(t_s^{k+1})$
deterministically. Also, using the first $\lambda$ bits of the packet,
the sensor can find the value of $\phi_{q(z(t_s))}$.  Similarly to
Proposition~\ref{lemmaoffeedback}, if the sensor has a causal
knowledge of the delay in the channel, it can calculate $\hat{x}(t)$
for all time~$t$.

\subsubsection{Sufficient packet size}\label{sec:suff-packet-size-complex}
Here we show that with a sufficiently large packet size, we can
achieve~\eqref{eq:jump-upp-HD} at all reception times $\{t^k_c\}_{k
  \in \mathbb{N}}$ using the quantization policy designed in
Section~\ref{sec:evencomplexdes}. The proof of the next result is in~Appendix~\ref{sec:proofs}.

\begin{figure}[htb]
	\centering
    \includegraphics[scale=0.3]{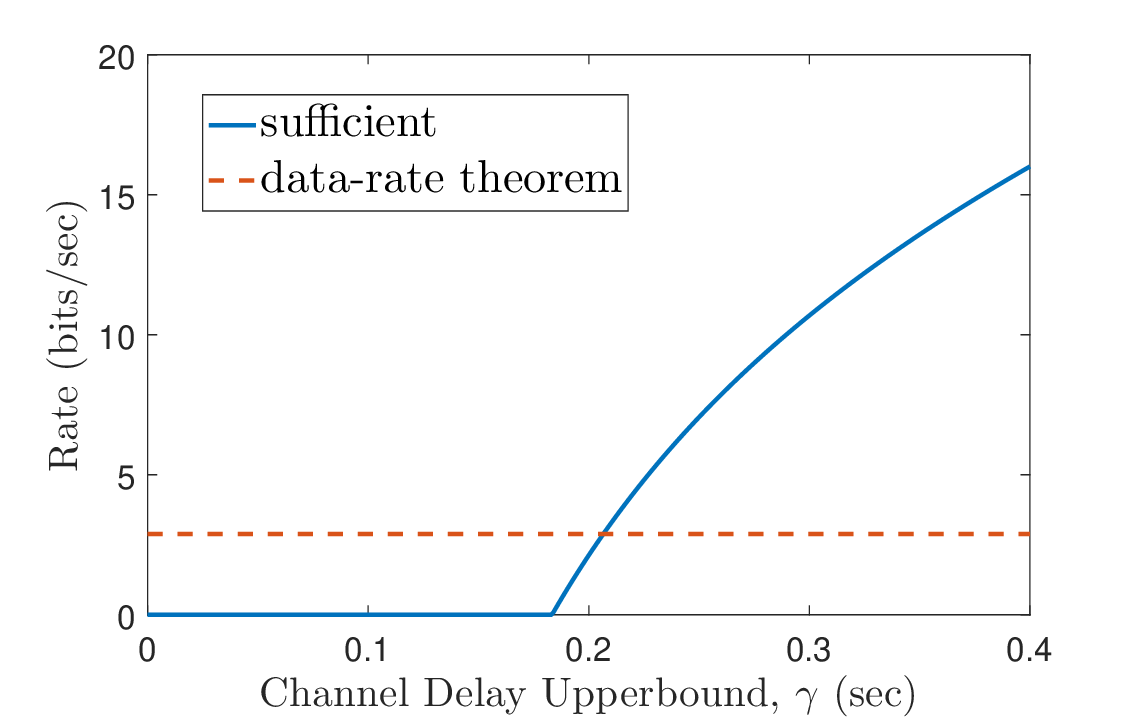} \\
	\caption{Sufficient information transmission rate~\eqref{inftranratecomplex} as a function of channel delay upper bound $\gamma$. Here $A=1+i$, $B=0.5$, $M=0.1$, $\rho_0=0.9$ and $b=1.0001$. Also ${\lambda=\log\big(\pi/{2\arcsin(\frac{7}{8})e^{\RE(A)\gamma}} \big)}$ and ${J=\frac{8M}{\RE(A)} \big( e^{\RE(A)\gamma}-1 \big) + 0.002}$. The rate dictated by the data-rate theorem (cf. Theorem~\ref{thm:necc-access-rate-HD}) is $2\RE(A)/\ln2=2.885$.}
    \label{fig:complex-rate-figures} 
	\vspace*{-2ex}
\end{figure}

\begin{theorem}\label{thm:suf-cond-complex} 
Consider the  model with plant dynamics~\eqref{sysconHD},
  estimator dynamics~\eqref{sysest}, triggering strategy~\eqref{eq:etsHHD}, and
  jump strategy~\eqref{eq:jumpst}. Assume $\|z(0)\|=\|x(0)-\hat{x}(0)\|<J$,
  then the quantization policy designed above achieves~\eqref{eq:jump-upp-HD}  for all reception times $\{t^k_c\}_{k \in \mathbb{N}}$ with any packet size lower bounded~by
\begin{align}\label{SufiHDpacket}
  &g(t_s) \ge \bar{g} \triangleq
  \\\nonumber
  &\max\left\{0,\lambda+\log \frac{\RE(A)b\gamma}{\ln \left(\frac{1+e^{-\RE(A)\gamma}\left(\rho_0 -\frac{M}{\RE(A)J}\left(e^{\RE(A)\gamma}-1\right)\right)}{2\sin(\pi/2^{\lambda+1})+1+\sqrt{2\zeta}}\right)}\right\},
  \end{align}
 provided $\cos\Big(\IM(A)\big(t_s-q(t_s)\big)\Big)=1-\zeta$
, $b>1$,
  \begin{subequations}
 \begin{align}\label{loweboundonrho012}
&\rho_0 \ge 
\\\nonumber
&\frac{M}{\RE(A)J}\left(e^{\RE(A)\gamma}-1\right)+
e^{\RE(A)\gamma}\left(2\sin(\pi/2^{\lambda+1})+\sqrt{2\zeta}\right),
\\
 \label{GG22}
& J\ge \frac{M}{\RE(A)\chi}\left(e^{\RE(A)\gamma}-1\right),
\hspace{0.4cm}
\sqrt{2\zeta}e^{\RE(A)\gamma}\le \chi',
\\
\label{GG33}
& \lambda>\log\left(\pi\Big/\arcsin\left(\frac{1-\chi-\chi'}{2e^{\RE(A)\gamma}}\right)
\right)-1,
\end{align}
\end{subequations}
where $0<\chi+\chi'<1$.
\end{theorem}
Combining the bound on the triggering rate from Remark~\ref{akhar12}
%
%
with Theorem~\ref{thm:suf-cond-complex}, it follows that  there exists an information transmission rate with 
\begin{align}\label{inftranratecomplex}
    R_s \le 
    \RE(A)\bar{g}\Big/\ln \left(\frac{J\RE(A)+M}{\rho_0J\RE(A)+M}\right),
\end{align}
that achieves~\eqref{eq:jump-upp-HD} for all reception times
$\{t^k_c\}_{k \in \mathbb{N}}$, and is therefore, sufficient to
ensure~\eqref{sysconHD} is ISpS.
\figref{fig:complex-rate-figures} shows the sufficient information
transmission rate in~\eqref{inftranratecomplex} as a function of the
delay upper bound~$\gamma$ on the channel delay. One can observe that
for small values of the delay, the sufficient information transmission
rate is smaller than the rate required by the data-rate result in
Theorem~\ref{thm:necc-access-rate-HD}. As $\gamma$ increases, the
sufficient information transmission rate increases accordingly.
{\blue{
\begin{remark}
{\rm 
Following the discussion of Remarks~\ref{ramrk3445} and~\ref{screm334}, when $\bar{g}=0$ in~\eqref{SufiHDpacket}, 
there is no need for
any data payload, 
and~\eqref{sysconHD} can be stabilized using only  timing information.
} \oprocend
\end{remark}
}}

\begin{remark}\label{akhar12!!22} {\rm Depending on whether the system
is real or complex, the corresponding  triggering criterion is based on the real or complex absolute value, resp., cf.~\eqref{eq:ets} and~\eqref{eq:etsHHD}. The controller needs to approximate the phase at which the state estimation error $z(t_s)$ hits the triggering radius. The real case is a particular case of our complex results, since the phase of $z(t_s)$ is then either $0$ or $\pi$. Thus, for the real case, in our sufficient design, only the first bits of the packet $p(t_s)$ denote the sign of $z(t_s)$. In the complex case, we devote the first $\lambda$ bits of the packet $p(t_s)$ for quantizing the phase of $z(t_s)$.
By putting $A=\RE(A)$, $\lambda=1$, and $\IM(A)=0$ (or $\zeta=0$), our sufficient condition for complex systems~\eqref{inftranratecomplex}, becomes~\eqref{inftranrate} except a factor $1+\sqrt{2}$, which makes~\eqref{inftranratecomplex} larger than~\eqref{inftranrate}. The reason for the difference is
the obtained 
upper bound in this case for the estimation error of the phase of $z(t_s)$ {\blue{(see~Eq.~\eqref{fjenieo111} in Appendix~\ref{sec:proofs})}}. In the real case,
the controller deduces $z(t_s)=J$ or $z(t_s)=-J$, and the estimation
error of the phase of $z(t_s)$ is zero.  }
  \oprocend
\end{remark}

\section{Conclusions}\label{sec:conc}
We have presented an event-triggered control scheme for the
stabilization of noisy, scalar real and complex, continuous, linear
time-invariant systems over a communication channel subject to random
bounded delay.  We have developed an algorithm for encoding-decoding
the quantized version of the estimated state, leading to the
characterization of a sufficient transmission rate for stabilizing
these systems. We also identified a necessary condition on the
transmission rate for real systems. Future work will study the
identification of necessary conditions on the transmission rate in
complex systems, develop event-triggered designs for vector
systems with real and complex eigenvalues, {\blue{and  the investigation of optimal values for the design parameters that balance the trade-offs between transmission rate and control performance.}}

{\bibliographystyle{IEEEtran}
\bibliography{mybib}
}


\begin{IEEEbiography}[{\includegraphics[width=1in,height=1.25in,clip,keepaspectratio]{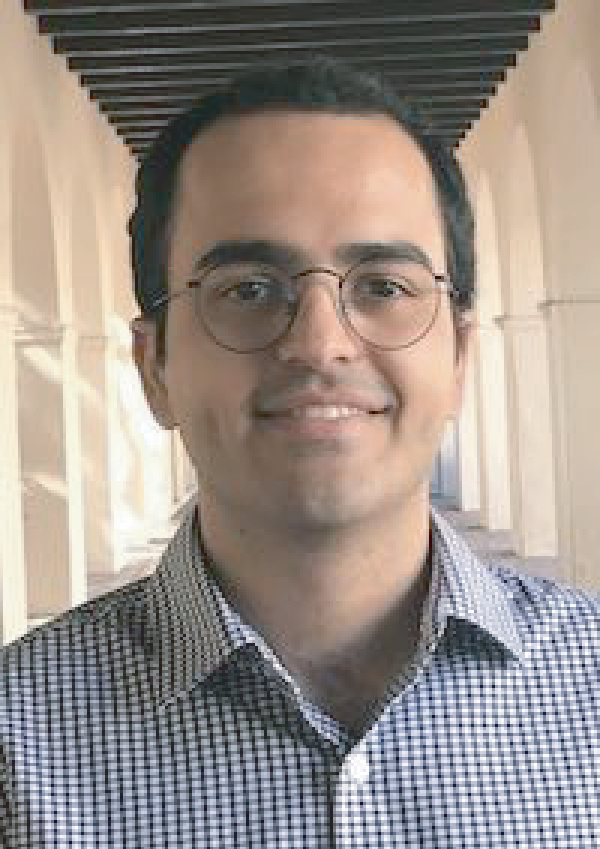}}]{Mohammad Javad Khojasteh}
  (S'14) did his undergraduate studies at Sharif University of
  Technology from which he received double-major B.Sc.\ degrees in
  Electrical Engineering and in Pure Mathematics, in 2015.  He
  received the M.Sc.\ and Ph.D.\ degrees in Electrical and Computer
  Engineering from University of California San Diego (UCSD), La
  Jolla, CA, in 2017, and 2019, respectively. Currently, he is a
  Postdoctoral Scholar in the Center for Autonomous Systems and Technologies (CAST) at California
  Institute of Technology, Pasadena, CA. 
\end{IEEEbiography}


\begin{IEEEbiography}[{\includegraphics[width=1in,height=1.25in,clip,keepaspectratio]{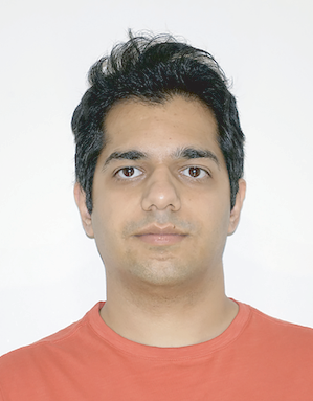}}]{Mojtaba Hedayatpour}
  received his bachelor's degree in aerospace engineering from Sharif
  University of Technology, Iran and his master's degree in industrial
  systems engineering from University of Regina, Canada. He has worked
  as Autonomous Systems Engineer at a startup company called Dot
  Technology Corporation in Canada where he developed guidance,
  navigation and control solutions for autonomous farming vehicles. He
  is currently the head of Artificial Intelligence team at the same
  company in Edmonton, Canada, where together with his team they
  develop autonomous solutions for farming applications.
\end{IEEEbiography}


\begin{IEEEbiography}[{\includegraphics[width=1in,height=1.25in,clip,keepaspectratio]{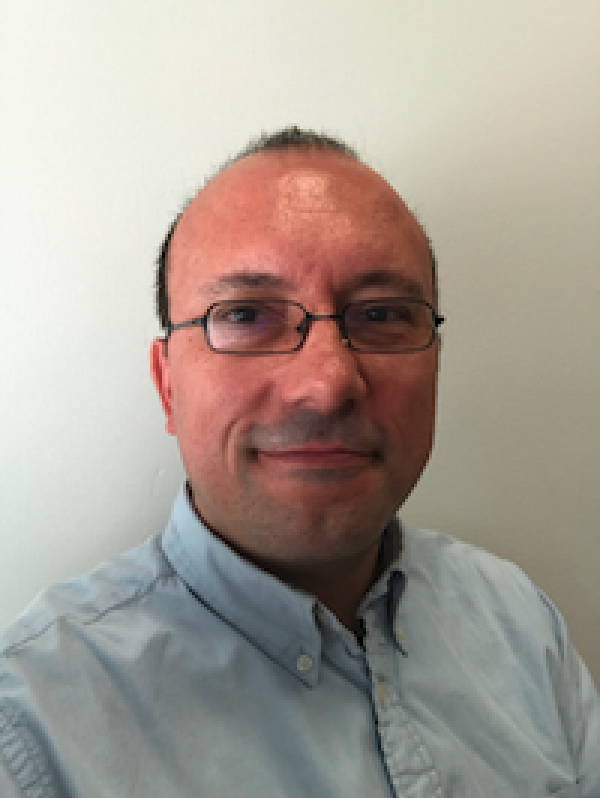}}]{Jorge Cort{\'e}s}
  (M'02-SM'06-F'14) received the Licenciatura degree in mathematics
  from Universidad de Zaragoza, Zaragoza, Spain, in 1997, and the
  Ph.D. degree in engineering mathematics from Universidad Carlos III
  de Madrid, Madrid, Spain, in 2001. He held postdoctoral positions
  with the University of Twente, Twente, The Netherlands, and the
  University of Illinois at Urbana-Champaign, Urbana, IL, USA. He was
  an Assistant Professor with the Department of Applied Mathematics
  and Statistics, University of California, Santa Cruz, CA, USA, from
  2004 to 2007. He is currently a Professor in the Department of
  Mechanical and Aerospace Engineering, University of California, San
  Diego, CA, USA. He is the author of Geometric, Control and Numerical
  Aspects of Nonholonomic Systems (Springer-Verlag, 2002) and
  co-author (together with F. Bullo and S. Mart{\'\i}nez) of Distributed
  Control of Robotic Networks (Princeton University Press, 2009). He is a Fellow of IEEE and
  SIAM. 
  At the IEEE Control Systems
  Society, he has been a Distinguished Lecturer (2010-2014), and is
  currently its Director of Operations and an elected member
  (2018-2020) of its Board of Governors. His current research
  interests include distributed control and optimization, network
  science, opportunistic state-triggered control and coordination,
  reasoning under uncertainty, and distributed decision making in
  power networks, robotics, and transportation.
\end{IEEEbiography}


\begin{IEEEbiography}[{\includegraphics[width=1in,height=1.25in,clip,keepaspectratio]{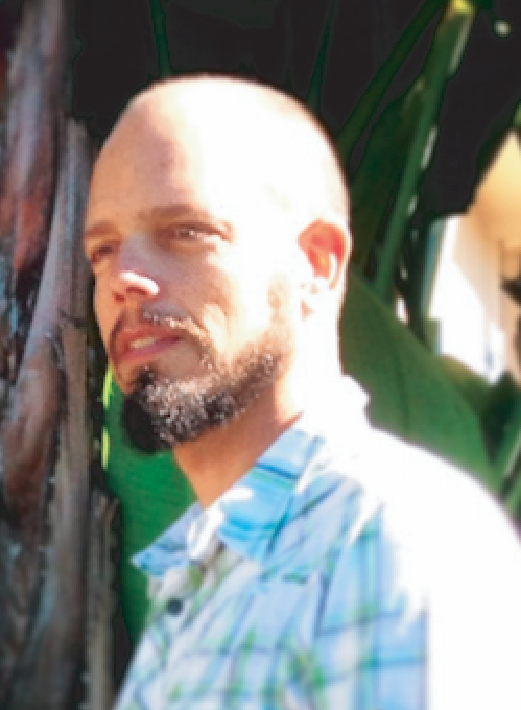}}]{Massimo Franceschetti} 
   received the Laurea degree
  (with highest honors) in computer engineering from the University  Federico II, Naples, Italy, in 1997, the M.S.\ and Ph.D.\ degrees in
  electrical engineering from the California Institute of Technology,
  in 1999, and 2003, respectively.  He is Professor of
  Electrical and Computer Engineering at the University of California
  at San Diego (UCSD). Before joining UCSD, he was a postdoctoral
  scholar at the University of California at Berkeley for two
  years. He has held visiting positions at the Vrije Universiteit
  Amsterdam, the \'{E}cole Polytechnique F\'{e}d\'{e}rale de Lausanne,
  and the University of Trento. His research interests are in physical
  and information-based foundations of communication and control
  systems. 
   He was awarded the C. H. Wilts
   Prize in 2003 for best doctoral thesis in electrical engineering at
   Caltech; the S.A.  Schelkunoff Award in 2005 for best paper in the
   IEEE Transactions on Antennas and Propagation, a National Science
   Foundation (NSF) CAREER award in 2006, an Office of Naval Research
  (ONR) Young Investigator Award in 2007, the IEEE Communications
   Society Best Tutorial Paper Award in 2010, and the IEEE Control
   theory society Ruberti young researcher award in 2012. 
  He has been elected fellow of the IEEE in 2018 and became a Guggenheim fellow for the natural sciences, engineering, in~2019.
\end{IEEEbiography}

\appendices

\section{Simulations}\label{sec:sim}
This section presents simulation results validating the proposed
event-triggered control scheme for real-valued plants (the interested
reader can find simulations for a complex-valued plant
in~\cite{cdc18paper}).
While our analysis is for continuous-time plants, we perform the simulations in discrete time with a small sampling time~$\delta'>0$. Thus, the minimum upper bound for the \magenta{communication network} delay is equal to two sampling times in the digital environment (this is because a delay of at most one sampling time might occur from the time that triggering occurs to the time that the sensor took a sample from the plant state and another delay of at most one sampling time might occur from the time that the packet is received to the time the control input is applied to the plant). 
We consider a linearized version of the two-dimensional problem of balancing an inverted pendulum mounted on a cart, where the motion of the pendulum is constrained in a plane and its position can be measured by an angle $\theta$.
The inverted pendulum has mass $m_1$, length $l$, and  moment of inertia $I$. Also, the pendulum is mounted on top of a cart of mass $m_2$, constrained to move in $y$ direction. 
The nonlinear equations governing the motion of the cart and pendulum are $(m_1+m_2)\ddot{y}+\nu\dot{y}+m_1l\ddot{\theta}\cos\theta-m_1l\dot{\theta}^2 \sin\theta=F$ and $(I+m_1l^2)\ddot{\theta}+m_1g_0lsin\theta=-m_1l\ddot{y}cos\theta$,
where $\nu$ is the damping coefficient between the pendulum and the cart and $g_0$ is the gravitational acceleration.
We define $\theta=\pi$ as the equilibrium position of the pendulum and $\phi$ as small deviations from~$\theta$. We derive the linearized equations of motion using small angle approximation, noting that this linearization is only valid for sufficiently small values of the delay upper bound~$\gamma$. Define the state variable $s=[y, \dot{y}, \phi, \dot{\phi}]^T$, where $y$ and $\dot{y}$ are the position and velocity of the cart respectively. Assuming $m_1=0.2$ kg, $m_2=0.5$ kg, $\nu=0.1$ N/m/s, $l=0.3$ m, $I=0.006$ kg/m$^2$, one can write the evolution of $s$ as
\begin{align}\label{sysconpendulum}
	\dot{s}=As(t)+Bu(t)+w(t),
\end{align}
where
\begin{align*}
\textstyle
    \label{AB_matrix}
    A =  \scriptstyle
    \begin{bmatrix}
     \scriptstyle 0 &  \scriptstyle 1 &  \scriptstyle 0 &  \scriptstyle 0 \\
     \scriptstyle 0 &  \scriptstyle  -0.1818 &  \scriptstyle 2.6730 &  \scriptstyle 0 \\
     \scriptstyle 0 &  \scriptstyle 0 &  \scriptstyle 0 &  \scriptstyle 1 \\
     \scriptstyle 0 & \scriptstyle -0.4545 & \scriptstyle 31.1800 &  \scriptstyle 0
    \end{bmatrix}
    \textstyle
    ,~B =  \scriptstyle
    \begin{bmatrix}
     \scriptstyle 0 \\
     \scriptstyle 1.8180 \\
     \scriptstyle 0 \\
     \scriptstyle 4.5450 
    \end{bmatrix}.
\end{align*}
In addition, we add the plant noise $w(t) \in \real^4$ to the linearized plant model, and  we assume that all of its elements are upper bounded by $M$. 
A simple feedback control law can be derived for
~\eqref{sysconpendulum} as $u=-Ks$, where $K = [ -1.00 \; -2.04 \;
20.36 \; 3.93 ]$.  is chosen such that $A-BK$ is Hurwitz.

The eigenvalues of the open-loop gain of the plant $A$ are $e = 
   [
    0 \; -5.6041 \; -0.1428 \; 5.5651
  ]
$. Thus, the open-loop gain of the plant $A$ is diagonalizable (all eigenvalues of $A$ are distinct).
Using the eigenvector matrix $P$,
we diagonalize the plant to obtain
\begin{equation}\label{AAmj}
\textstyle
    \dot{\tilde{s}} = \tilde{A}\tilde{s}(t)+\tilde{B}\tilde{u}(t)+\tilde{w}(t),
\end{equation}
where
\begin{align*}
\textstyle
    \tilde{A} = \scriptstyle
    \begin{bmatrix}
    \scriptstyle 0 & \scriptstyle 0 & \scriptstyle  0 & \scriptstyle 0 \\
    \scriptstyle 0 & \scriptstyle -5.6041 & \scriptstyle 0 & \scriptstyle 0 \\
    \scriptstyle 0 & \scriptstyle 0 & \scriptstyle -0.1428 & \scriptstyle 0 \\
    \scriptstyle 0 & \scriptstyle 0 & \scriptstyle 0 & \scriptstyle 5.5651
    \end{bmatrix}
    \textstyle
    ,~\tilde{B} = \scriptstyle
    \begin{bmatrix}
    \scriptstyle 10.0000 \\
    \scriptstyle -2.3865 \\
    \scriptstyle 10.0979 \\
    \scriptstyle 2.2513 
    \end{bmatrix},
\end{align*}
where $\tilde{s}(t) = P^{-1}s(t)$
and $\tilde{w}(t) = P^{-1}w(t)$. Also, $\tilde{u}(t)=-\tilde{K}\tilde{s}(t)$ where $\tilde{K}=KP$.

For the first three coordinates of the diagonalized plant
in~\eqref{AAmj} the state estimation
$\hat{s}$ at the controller simply constructs as $\dot{\hat{s}}_i = \tilde{A}_i\hat{s}(t)+\tilde{B}_i\tilde{u}(t)$,
starting from $\hat{s}_i(0)$ for $i \in \{1,2,3\}$, where $\tilde{A}_i$ and $\tilde{B}_i$ denote the $i^{th}$ row of $\tilde{A}$ and $\tilde{B}$. Since the first three eigenvalues of $A$ are non-positive, they are inherently stable. Thus, by the data theorem~\cite{sharon2012input} there is no need to use the communication \magenta{network} for them, and since $\tilde{A}-\tilde{B}\tilde{K}$ is Hurwitz, $\tilde{u}(t)=-\tilde{K}\tilde{s}(t)$ renders them ISS with respect to system disturbances.
%
Now we apply Theorem~\ref{thm:suf-cond-ET} to the fourth mode of the plant, which is unstable, to make the whole plant ISpS. In fact, we use the packet size given in~\eqref{efkfek!!!45hgree} for the simulations.
Using the problem formulation in Section~\ref{sec:setup},
the estimated state for the unstable mode $\hat{s}_4$ evolves during
the inter-reception times as
\begin{align}\label{esunstablemode}
  \dot{\hat{s}}_4(t)= 5.5651\hat{s}_4(t)+2.2513\tilde{u}(t), \quad t
  \in (t_c^k,t_c^{k+1}),
\end{align}
starting from $\hat{s}_4(t_c^{k+})$ and $\hat{s}_4(0)$.
Also, a triggering occurs when $ |\tilde{z}_4(t)|=|\tilde{s}_4(t)-\hat{s}_4(t)| = J$,
where $|\tilde{z}_4(t)|$ is the estate estimation error for the
unstable mode, and assuming the previous packet is already delivered to the controller. In the simulation environment, 
since the sampling time is small, a triggering happens as soon as $|\tilde{z}_4(t)|$ is equal or greater than $J$ and the previous packet has been received by the controller.
Let $\lambda_4= 5.5651$ be the eigenvalue corresponding
to the unstable mode. By Theorem~\ref{thm:suf-cond-ET}, we choose $J=(M/(\lambda_4\rho_0))(e^{\lambda_4\gamma}-1)+0.005$,
and the size of the packet for all $t_s$ to be~\eqref{efkfek!!!45hgree},
where $b=1.0001$ and $\rho_0=0.9$.
\begin{figure*}[t]
  \centering
  \begin{tabular}{c c c c}
    \subfigure[]{\includegraphics[width=.24\linewidth]{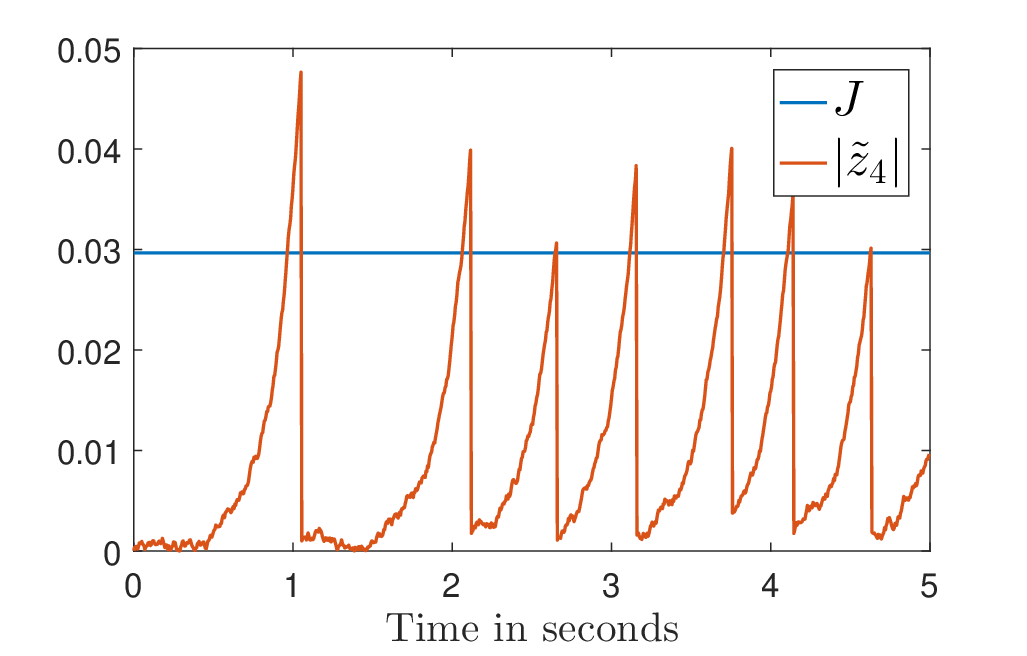}} 
   \subfigure[]{ \includegraphics[width=.24\linewidth]{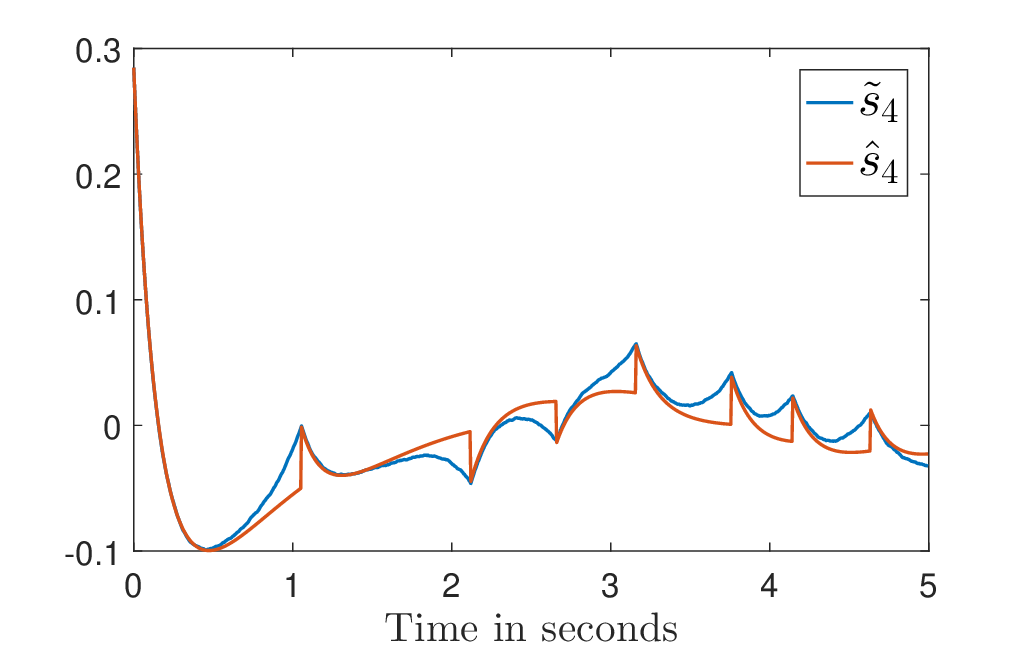}}
    \subfigure[]{\includegraphics[width=.24\linewidth]{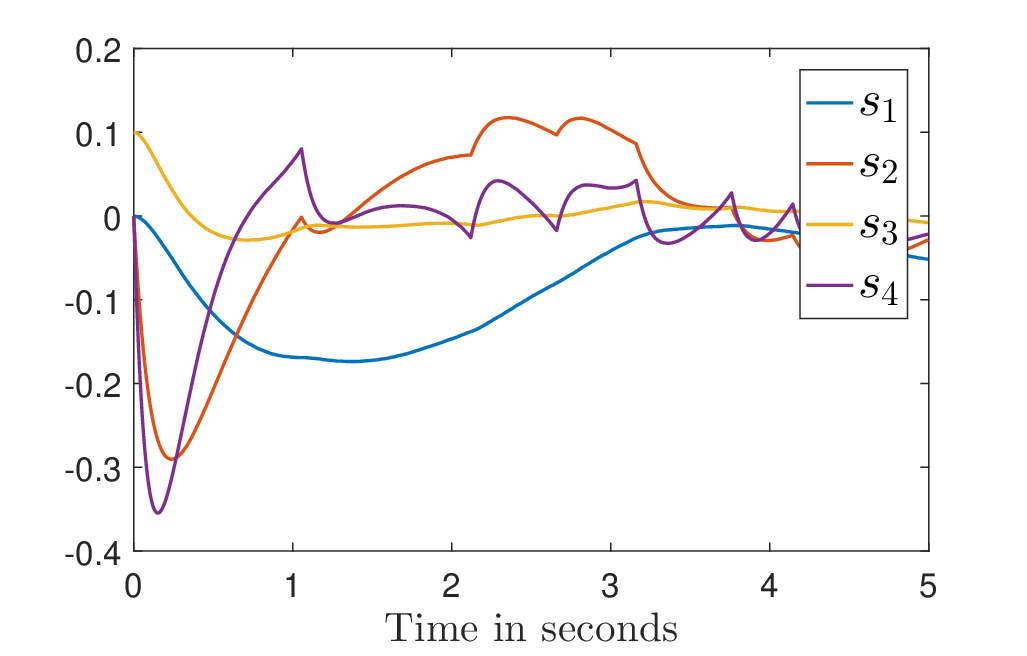}}
    \subfigure[]{\includegraphics[width=.24\linewidth]{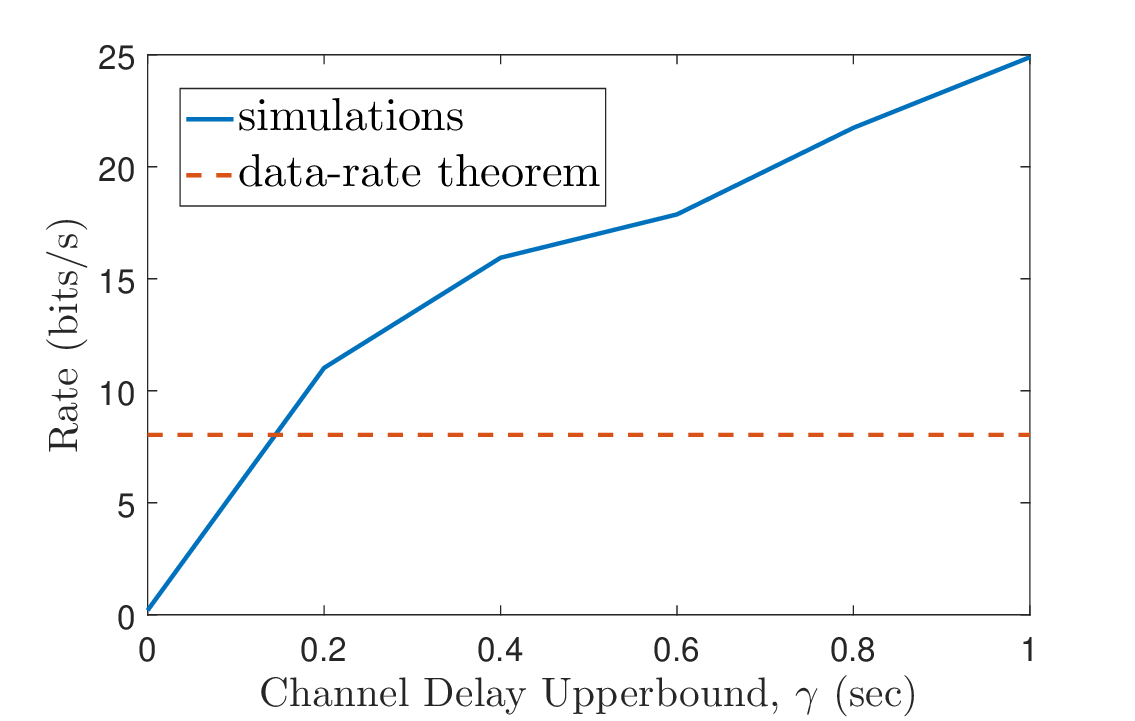} }
  \end{tabular}
\caption{Simulation results for the linearized inverted pendulum on a cart example. (a) shows the evolution of the absolute value of the state estimation error (a) for the unstable mode of the plant in~\eqref{AAmj}. (b) shows the evolution of the unstable state in~\eqref{AAmj} and its estimate in~\eqref{esunstablemode}. (c) shows the evolution of all the states in~\eqref{sysconpendulum}. (d) shows the information transmission rate in the simulation as compared to the data-rate theorem. Note that the rate does not start at $\gamma=0$ because the minimum channel delay upper bound is equal to two sampling times (0.005 seconds in this example). 
The simulation parameters are $\tilde{s}(0)={P^{-1}}[0,0,0,0.1001]^T$, $\hat{s}(0)={P^{-1}}[0,0,0,0.10]^T$, simulation time $T=5$ seconds, and sampling time $\delta'=0.005$ seconds,   For (a)-(c), $\gamma=0.1$ sec, $g(t_s)=4$ bits, $M=0.05$, and in (d) $g(t_s)$ is calculated using~\eqref{efkfek!!!45hgree} with $M=0.2$. 
}
\label{simrespand}
	\vspace*{-2ex}
\end{figure*}

\begin{figure*}[h]
  \centering
  \begin{tabular}{c c c}
    \subfigure[]{\includegraphics[width=.33\linewidth]{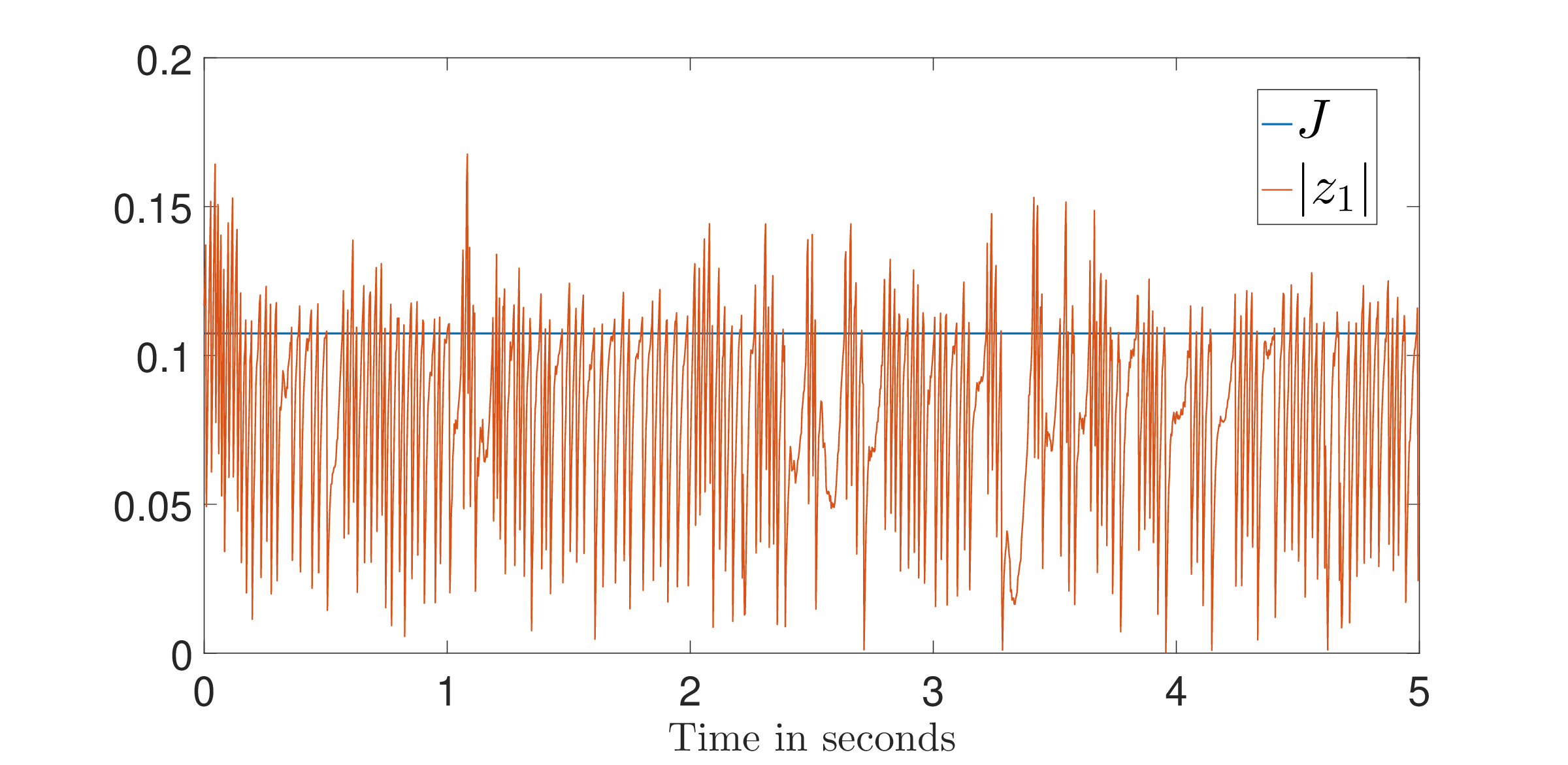}} 
    \subfigure[]{ \includegraphics[width=.33\linewidth]{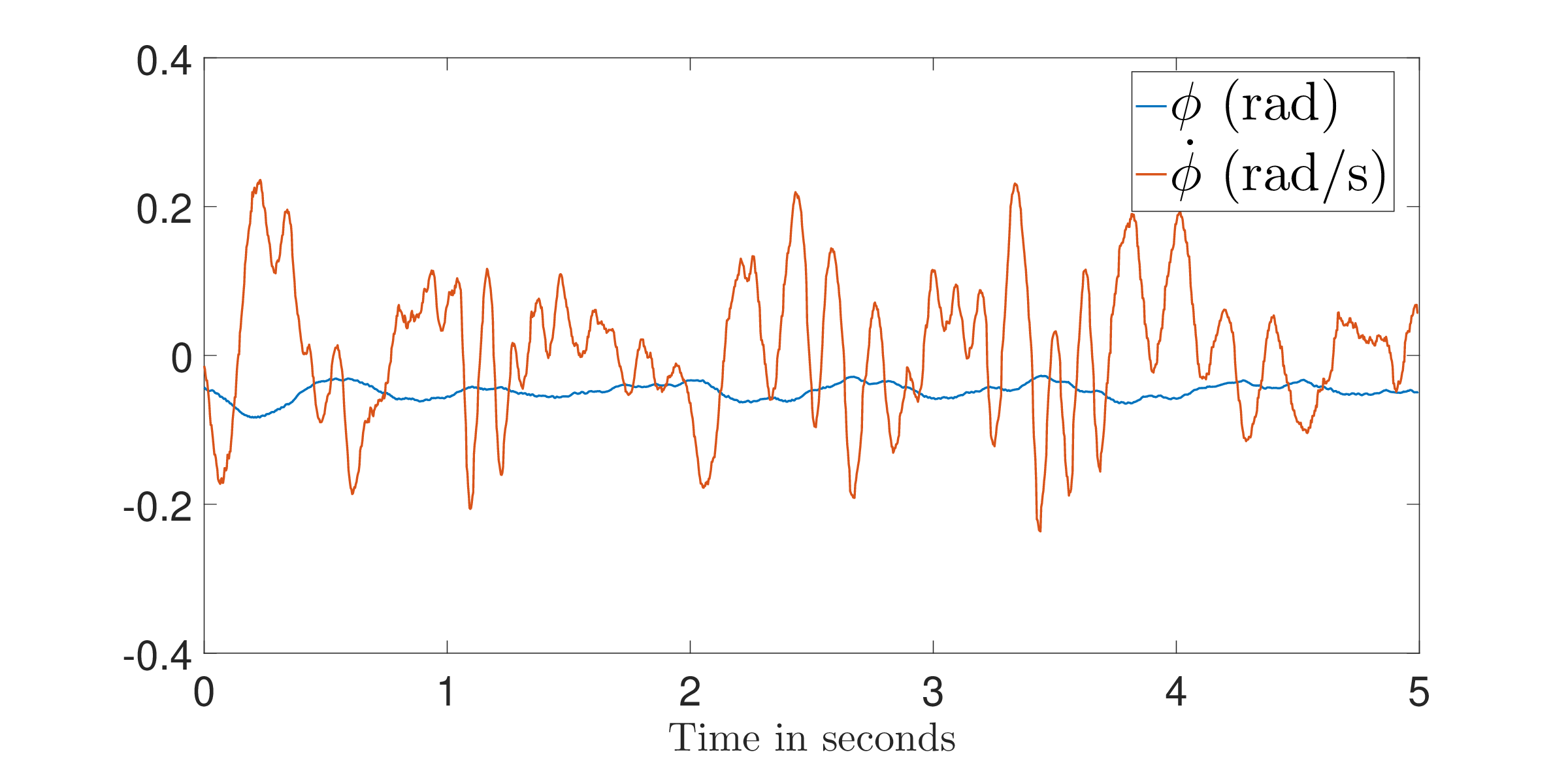}}
    \subfigure[]{\includegraphics[width=.33\linewidth]{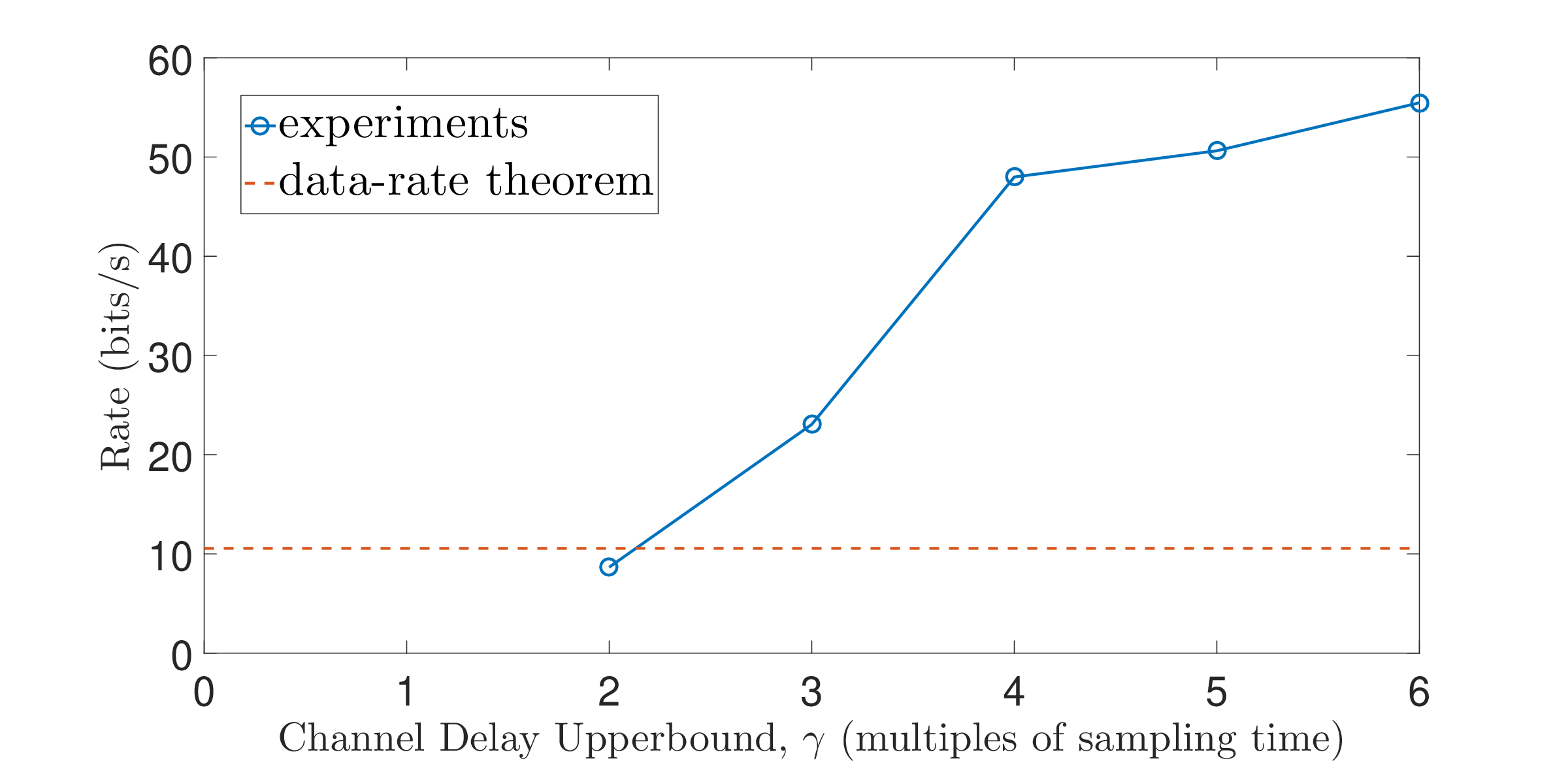}}
  \end{tabular}
\caption{
Experimental results for controlling an inverted pendulum with the proposed event-triggered control strategy. (a) shows the triggering threshold $J$ and the estimation error in the pendulum's angular position $z_1=\phi-\hat{\phi}$, where $\phi$ is the sensor measurement and $\hat{\phi}$ is the estimate of the angular position. (b) shows the actual angular position and velocity of the pendulum, with the former staying close to zero degrees (the desired upright position). In (a) and (b) the delay upper bound is set to five sampling times of the system (which is equal to 0.015 seconds) and the packet size is found to be 7 bits. (c) shows the information transmission rate in the experiments compared with the entropy rate of the system. The rate calculated from the experiments does not start at zero because the minimum channel delay upper bound is equal to two sampling times (0.006 seconds). The entropy rate of the system is 10.56 bits/sec, while the minimum transmission rate for delay bound equal to two sampling times is 8.66 bits/sec. 
}
\label{fig:exp_results}
\end{figure*}

\figref{simrespand}(a) shows the triggering threshold for $\tilde{s}_4$ in~\eqref{AAmj} and the absolute value of the state estimation error for the unstable coordinate, that is, $|\tilde{z}_4(t)|=|\tilde{s}_4(t)-\hat{s}_4(t)|$. As soon as the absolute value of this error is equal or greater than the triggering threshold, the sensor transmits a packet, and the jumping strategy adjusts $\hat{s}_4$ at the reception time to 
ensure the plant is ISpS. 
Note that the amount this error exceeds the triggering threshold depends on the random \magenta{communication network} delay upper bounded by $\gamma$. 
\figref{simrespand}(b) presents the evolution of the unstable state in~\eqref{AAmj} and its estimation in~\eqref{esunstablemode}. \figref{simrespand}(c) shows the evolution of all the actual states of the  linearized plant~\eqref{sysconpendulum}. 
Finally,~\figref{simrespand}(d) presents the simulation of
information transmission rate versus the delay upper bound $\gamma$ in
the communication \magenta{network} for stabilizing the linearized
model of the inverted pendulum.  For small $\gamma$, the plant is ISpS
with an information transmission rate smaller than the one prescribed
by the data-rate theorem.

\begin{figure}[tbh]
   \centering
   \includegraphics[scale=1]{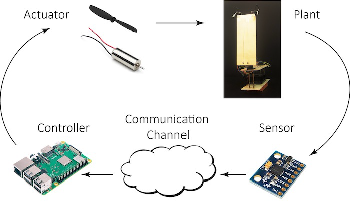} 
   \caption{Architecture and components of the prototype. }
   \label{fig:sys-components}
\end{figure}

\begin{remark}
\label{tam733}
{\rm \magenta{For further validation, we have also
      experimentally implemented the proposed event-triggered control
      strategy on an inverted pendulum controlled by two
      propellers as shown in \figref{fig:sys-components}.
      The robot used for the experiments is built using off-the-shelf
      components.  Specifically, the frame is built with plywood
      sheets, we employ an InvenSense MPU6050 MEMS sensor (which
      consists of a 3-axis accelerometer and a 3-axis gyroscope),
      using a complementary filter to estimate the pendulum's angle
      and angular velocity, and we have a Raspberry Pi Model~3 acting
      as the computation unit as well as the controller.  Finally, two
      small DC motors equipped with two identical propellers are used
      as actuators. \figref{fig:exp_results}(a) shows the
      evolution of the pendulum angle estimation error $z_1$ in time
      and~\figref{fig:exp_results}(b) shows the angular position and velocity of the
      pendulum, where zero angle represents the upright
      position of the pendulum. We also ran a second set of
      experiments and calculated the information transmission rates
      using~\eqref{efkfek!!!45hgree} as a function of the delay upper
      bound, cf. \figref{fig:exp_results}(c). The reason for the larger number of jumps in the experiments compared to the simulation is due to the additional uncertainty introduced by the nonlinear behavior of the system. Nevertheless, the same qualitative phase transition behavior is observed in \figref{simrespand}(d) and \figref{fig:exp_results}(c).  The interested reader is
      referred to~\cite{cdc19paper} for further details of these
      experiments and validation\footnote{{\blue{The code can also be found at 
      
      \texttt{https://github.com/mkhojas/Event-Triggered-Firmware}.}}}}.
      }
  \oprocend
\end{remark}

{\blue{
\begin{remark}
Several plots and discussions that illustrate the dependency of  our sufficient~\eqref{inftranrate} and necessary~\eqref{necessT} rates on the plant disturbances $M$ and the design parameter $J$, along with plots and discussions that illustrate the effect of design parameters $b$ and $\rho_0$  on the sufficient rate~\eqref{inftranrate} are available in~Appendix~\ref{sec:design-parameters}. \oprocend
\end{remark}
}}

\newpage
\onecolumn
\newpage
\twocolumn

\section{Proofs}\label{sec:proofs}
\begin{IEEEproof}[Proof of Proposition~\ref{lemmaoffeedback}]
{\blue{
Note that we are using the proposed quantizer in~\figref{encoder}, hence given $t_s^{k}$, $q(t_s^{k})$ gets identified deterministically. Therefore, given $t_c^k$ (which can be calculated using the causal knowledge of $k^{\text{th}}$ communication delay) and~\eqref{ctrlapp1}, the sensor constructs the value of $z(t_c^{k+})$ and determines the value of $\hat{x}(t_c^{k+})$.}}
  %
  %
\end{IEEEproof}

\begin{IEEEproof}[Proof of Lemma~\ref{NecesaaQuanClas}]
Without loss of generality assume that $z(t_s)=J$ throughout this proof. We also consider the realization of $w(t)=M$ for all time $t$.
We first show $\beta$ is the time needed for the state estimation error to
grow from $z(t_s)$ to $z(t_s)+2 J$. From~\eqref{solz}, we deduce at delay $\beta$,
\begin{align}\label{JasonM}
z(t_c)=e^{A\beta}J+(M/A) \left(e^{A\beta}-1\right).
\end{align}
By combining~\eqref{JasonM}, the bound on $\beta$, and $z(t_s)=J$ it follows 
$z(t_c)=z(t_s)+2J$. Hence, the value of $z(t_c)$ sweeps an area of measure $2J$ when the delay takes values in
$[0,\beta]$.

We continue by distinguishing between two classes of quantization cells.
We call a quantization cell \textit{perfect}, if its measure is equal to $2 J$, and when the measure of a quantization cell is less than  $2J$ we call it \textit{defective}. Using these definitions we now prove the occurrence of~\eqref{eq:deltaz} with  delay of at most $\beta$, in three different cases.
First, when $z(t_s)$ is in a perfect
cell, clearly for a  delay of at most $\beta$ we have 
$|z(t_c^k)-\bar{z}(t_c^k)|\geq J$, and~\eqref{eq:deltaz} follows. Second, when $z(t_s)$ is in a defective 
cell which is adjacent to a perfect 
cell, for a delay of at most $\beta$  the value of $z(t_c)$ sweeps the area of the defective cell and $z(t_c)$ enters the adjacent perfect 
cell.
Thus, with a delay at most $\beta$ we have $|z(t_c^k)-\bar{z}(t_c^k)|\geq J/2$, where $\bar{z}(t_c^k)$ is the center of the adjacent perfect 
cell with radius $J$, and~\eqref{eq:deltaz} follows. 
It remains to check the assertion when $z(t_s)$ is in a defective quantization cell, which is adjacent to another defective quantization cell.
Due to the restriction on the quantization policies as in  Assumption~\ref{Defnition11}, the sensor transmits the minimum required bits to divide the uncertainty set at the controller to quantization cell of measure of at most $2J$. 
If the measure of union of two adjacent 
cells is at most $2J$, these two balls could be replaced by one quantization cell 
to reduce the number of quantization cells.
As a consequence, under Assumption~\ref{Defnition11}, the measure of union of two adjacent quantization cells is greater than $2J$. 
Assume the defective quantization cell that contains $z(t_s)$ is of the measure $\mu_1$, and the measure of the adjacent defective 
cell is $\mu_2$. As a result, we have $\mu_1+\mu_2>2J$. Therefore, at least one of the $\mu_1$ or $\mu_2$ is at least $J$, thus with a delay of at most $\beta$, we have $|z(t_c^k)-\bar{z}(t_c^k)|\geq J/2$, and~\eqref{eq:deltaz} follows.  
\end{IEEEproof}

\begin{IEEEproof}[Proof of Theorem~\ref{thm:necc-access-rate-HD}]
  It is enough to prove the assertion when $w(t)=0$.  
  By rewriting~\eqref{sysconHD} when $w(t)=0$,
  $\dot{\RE(x)}+i\dot{\IM(x)}=\RE(A)\RE(x)-\IM(A)\IM(x)+i(\RE(A)\IM(x)+\IM(A)\RE(X))$,
  which is equivalent to
  \begin{align*}
  \scriptstyle
    \begin{bmatrix}
    \scriptstyle \dot{\RE(x)} \\ \scriptstyle \dot{\IM(x)}
    \end{bmatrix}
    =
   \begin{bmatrix}
    \scriptstyle \RE(A) & \scriptstyle -\IM(A) \\\scriptstyle \IM(A) & \scriptstyle \RE(A)
    \end{bmatrix}
  \begin{bmatrix}
    \scriptstyle \RE(x)(t) \\\scriptstyle \IM(x)(t)
    \end{bmatrix}
    .
  \end{align*}
  Since $\|x\|=\sqrt[]{\RE(x)^2+\IM(x)^2}$, if $\RE(x)$ or $\IM(x)$ becomes unbounded, $\|x\|$ becomes unbounded. Consequently, using~\cite[Theorem~1]{hespanha2002towards}, we need to have 
  \begin{align*}
    \textstyle
    R_c\ge \Tr\left( \begin{bmatrix}
    \scriptstyle 
    \RE(A) & \scriptstyle  -\IM(A) \\ \scriptstyle \IM(A) & \scriptstyle \RE(A)
    \end{bmatrix}\right)/\ln 2.~~~~~~
 \end{align*}
\end{IEEEproof}

\begin{IEEEproof}[Proof of Theorem~\ref{thm:suf-cond-complex}]
In our design, the controller estimates $z(t_c)$ as in~\eqref{zbar-tc-estimate}, and the encoding-decoding scheme is as depicted in~\figref{encoder} and~\ref{fig:triggering-cricle}. Using~\eqref{solz},~\eqref{zbar-tc-estimate}, and the triangle inequality,
\begin{align}\label{Shouro2}
&\|z(t_c)-\bar{z}(t_c)\|\le 
\\\nonumber
&\left\|\left(e^{A(t_c-t_s)}z(t_s)-e^{A(t_c-q(t_s))}q\left(z(t_s)\right)\right)\right\|
\\\nonumber
&+\left\|\int_{t_s}^{t_c} e^{A(t_c-\tau)}w(\tau) d\tau\right\|.
\end{align}
Similarly to~\eqref{calculation}, since $\|w(t)\|\le M$, the second summand in~\eqref{Shouro2} is upper bounded as
\begin{align}\label{Shouro4}
\left\|\int_{t_s}^{t_c} e^{A(t_c-\tau)}w(\tau) d\tau\right\|\le
 \frac{M}{\RE(A)}\left(e^{\RE(A)\gamma}-1\right).
\end{align}
To find a proper upper bound on the first summand in~\eqref{Shouro2}, assuming $q\left(z(t_s)\right)=z(t_s)-v_1$ and $q(t_s)=t_s-v_2$,
\begin{align}\label{Shouro6}
&\left\|e^{At_c}\left(e^{-At_s}z(t_s)-e^{Aq(t_s)}q\left(z(t_s)\right)\right)\right\|=
\\\nonumber
&\left\|e^{A(t_c-t_s)}\left(z(t_s)-e^{Av_2}\left(z(t_s)-v_1\right)\right)\right\|\le 
\\ \nonumber
&e^{\RE(A)\gamma}\left(J\|1-e^{Av_2}\|+e^{\RE(A)v_2}\left\|v_1\right\|\right).
\end{align}

Next, we find an upper bound of $\|v_1\|$. Since the sensor devotes $\lambda$ bits to transmit a quantized version of the phase of $z(t_s)$ to the controller, we have the upper bound $|\omega|\le \pi/2^{\lambda}$ on the difference of the phases of $z(t_s)$ and  $q(z(t_s))$.  Also, over $[-\pi,\pi]$, the cosine function is concave, with global maximum at $0$. Hence,  as depicted in~\figref{fig:triggering-cricle}, from the law of cosines,
\begin{align}
\label{fjenieo111}
&\|v_1\|=\|z(t_s)-q\left(z(t_s)\right)\|\le
\\\nonumber
&\sqrt[]{2J^2(1-\cos(\pi/2^{\lambda}))}=
2J\sin(\pi/2^{\lambda+1}).
\end{align}

Combining this with~\eqref{Shouro6}, the first summand in~\eqref{Shouro2} is upper bounded by
\begin{align*}
Je^{\RE(A)\gamma}\left(\|1-e^{Av_2}\|+2e^{\RE(A)v_2}\sin(\pi/2^{\lambda+1})\right).
\end{align*}
Note that 
$\|1-e^{Av_2}\|^2=(1-e^{\RE(A)v_2})^2+2e^{\RE(A)v_2}\zeta$,
where
    $\cos(\IM(A)v_2)=1-\zeta$, and $0\le\zeta\le2$.
Thus, the first summand in~\eqref{Shouro2} is upper bounded by
\begin{align*}
Je^{\RE(A)\gamma}\Big(|1-e^{\RE(A)v_2}|+\sqrt{2e^{\RE(A)v_2}\zeta}+
\\
2e^{\RE(A)v_2}\sin(\pi/2^{\lambda+1})\Big).
\end{align*}
For any positive real number $\epsilon$ we know $\epsilon+1/\epsilon \ge 2$, hence,
$e^{\RE(A)v_2}-1\ge 1-e^{-\RE(A)v_2}$.
Therefore, for the rest of the proof, and without loss of generality, we assume $v_2\ge 0$, and the first summand in~\eqref{Shouro2} is upper bounded by
\begin{align}\label{so135w22}
Je^{\RE(A)\gamma}\Big(e^{\RE(A)v_2}-1+\sqrt{2\zeta}e^{\RE(A)v_2}+
\\\nonumber
2e^{\RE(A)v_2}\sin(\pi/2^{\lambda+1})\Big).
\end{align}
Combining~\eqref{Shouro2},~\eqref{Shouro4}, and~\eqref{so135w22},
\begin{align}\label{MDF77}
&e^{\RE(A)v_2}\le
\\\nonumber
& \frac{1+e^{-\RE(A)\gamma}\left(\rho_0 -\frac{M}{\RE(A)J}\left(e^{\RE(A)\gamma}-1\right)\right)}{2\sin(\pi/2^{\lambda+1})+1+\sqrt{2\zeta}}
\end{align}
which suffices to ensure~\eqref{eq:jump-upp-HD}.
%
%
%
Recalling $v_2=t_s-q(t_s)$, using~\eqref{Quntrulesuff1C} and by setting
\begin{align*}
&\frac{b\gamma}{2^{g(t_s)-\lambda}}\leq 
\\\nonumber
&\frac{1}{\RE(A)}\ln \left(\frac{1+e^{-\RE(A)\gamma}\left(\rho_0 -\frac{M}{\RE(A)J}\left(e^{\RE(A)\gamma}-1\right)\right)}{2\sin(\pi/2^{\lambda+1})+1+\sqrt{2\zeta}}\right),
\end{align*}
%
\eqref{MDF77} is ensured.
Hence, the packet size in~\eqref{SufiHDpacket} is sufficient to ensure~\eqref{eq:jump-upp-HD} 
for all reception times.
However,~\eqref{MDF77} is well defined only when the upper bound   is at least one, namely
\begin{align*}
e^{-\RE(A)\gamma}\left(\rho_0 -\frac{M}{\RE(A)J}\left(e^{\RE(A)\gamma}-1\right)\right)
\ge
\\\nonumber
2\sin(\pi/2^{\lambda+1})+\sqrt{2\zeta},
\end{align*}
which holds because of~\eqref{loweboundonrho012}. Moreover, the design parameter $\rho_0$ in~\eqref{eq:jump-upp-HD} should be in the open interval $(0,1)$. Therefore, the lower bound in~\eqref{loweboundonrho012} should be smaller than $1$, namely
 \begin{align*}
\frac{M}{\RE(A)J}\left(e^{\RE(A)\gamma}-1\right)+e^{\RE(A)\gamma}(2\sin(\pi/2^{\lambda+1})+\sqrt{2\zeta})<1.
\end{align*}
The result now follows by noting that~\eqref{GG22}, and~\eqref{GG33} ensure this inequality holds. 
\end{IEEEproof}

{\blue{
\section{Dependency of sufficient and necessary conditions on design parameters}\label{sec:design-parameters}


Here we discuss the dependency of the sufficient~\eqref{inftranrate} and necessary~\eqref{necessT} rates on the plant disturbance bound~$M$ and the triggering threshold~$J$, and also examine the effect of design parameters $b$ and $\rho_0$ on the sufficient rate~\eqref{inftranrate}.

\subsubsection{Plant disturbance upper bound~$M$}
\figref{fig:graphs-M} illustrates the effect of  plant disturbance upper bound $M$ on the sufficient~\eqref{inftranrate} and necessary~\eqref{necessT} rates.
The results that are shown in this figure demonstrate that as $M$ increases, because of the increased uncertainty in the state estimation, the information transmission rate required
for~\eqref{syscon} to be ISpS increases as well. 

\begin{figure}[htb!]
  \centering
  \subfigure[]{
  \includegraphics[scale=0.40]{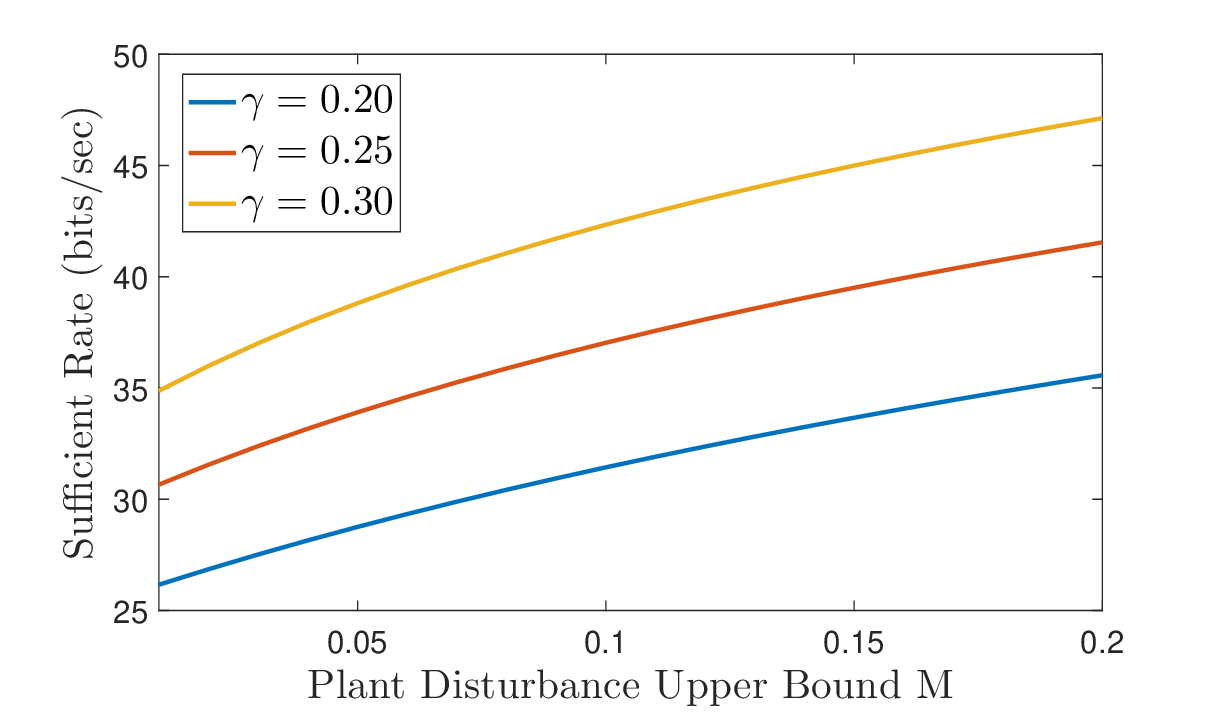}}
  \subfigure[]{
  \includegraphics[scale=0.40]{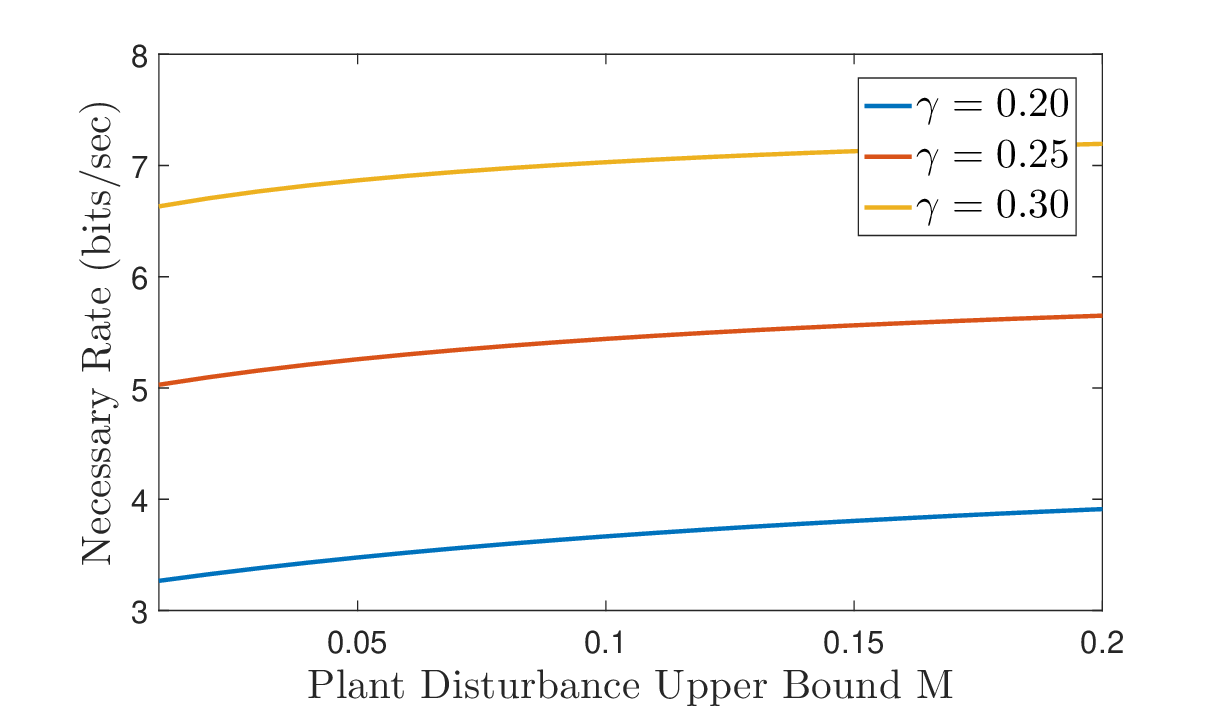}}
  \caption{(a) Sufficient and (b) necessary transmission rate vs plant disturbance upper bound $M$. Other simulation parameters are as follows: $A=5.5651$, $b=1.01$, $\rho_0=0.1$, and $J=(M/(A\rho_0))(e^{A\gamma}-1)+0.2$.}
  \label{fig:graphs-M}
\end{figure}

\subsubsection{Triggering threshold $J$}
\label{TTLSECr33}
\figref{fig:graphs-J} illustrates the effect of the triggering threshold $J$ on the sufficient~\eqref{inftranrate} and necessary~\eqref{necessT} rates. A smaller $J$  means more triggering, leading to closer agreement with the continuous-time dynamics. In the absence of disturbances, $J$ can go all the way down to zero to exactly match the continuous-time performance. However, in our paper, due to the presence of the disturbance, we have a lower bound on how small $J$ can be. In fact, to derive the necessary condition we assumed $J \ge M/A$ (cf. Lemma~\ref{inclusionsets} and Theorem~\ref{THM:NECESAAPP}), and our sufficient condition holds provided that $J>\frac{M}{A\rho_0}(e^{A\gamma}-1)$ (cf. Theorem~\ref{thm:suf-cond-ET}). In~\figref{fig:graphs-J}, we display the rates for $J - \frac{M}{A\rho_0}(e^{A\gamma}-1) \in [0.01,0.20]$.

\begin{figure}[htb]
  \centering
    \subfigure[]{
  \includegraphics[scale=0.40]{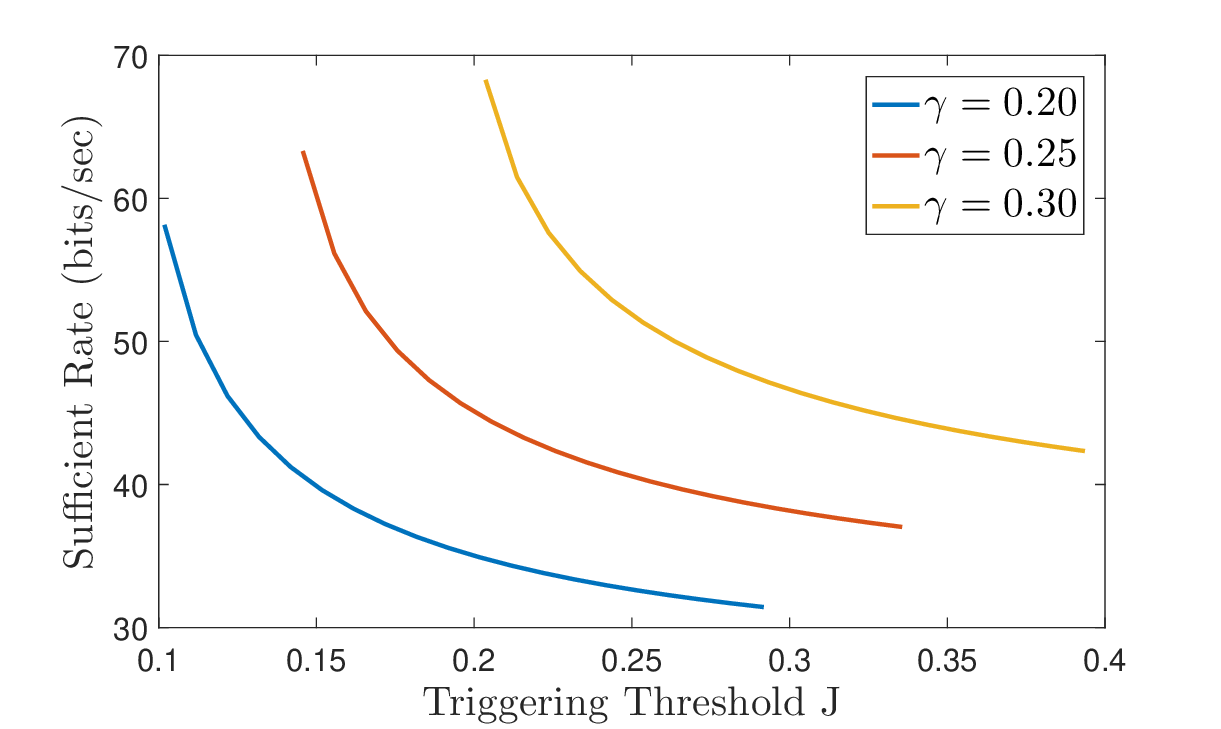}}
  \subfigure[]{
    \includegraphics[scale=0.40]{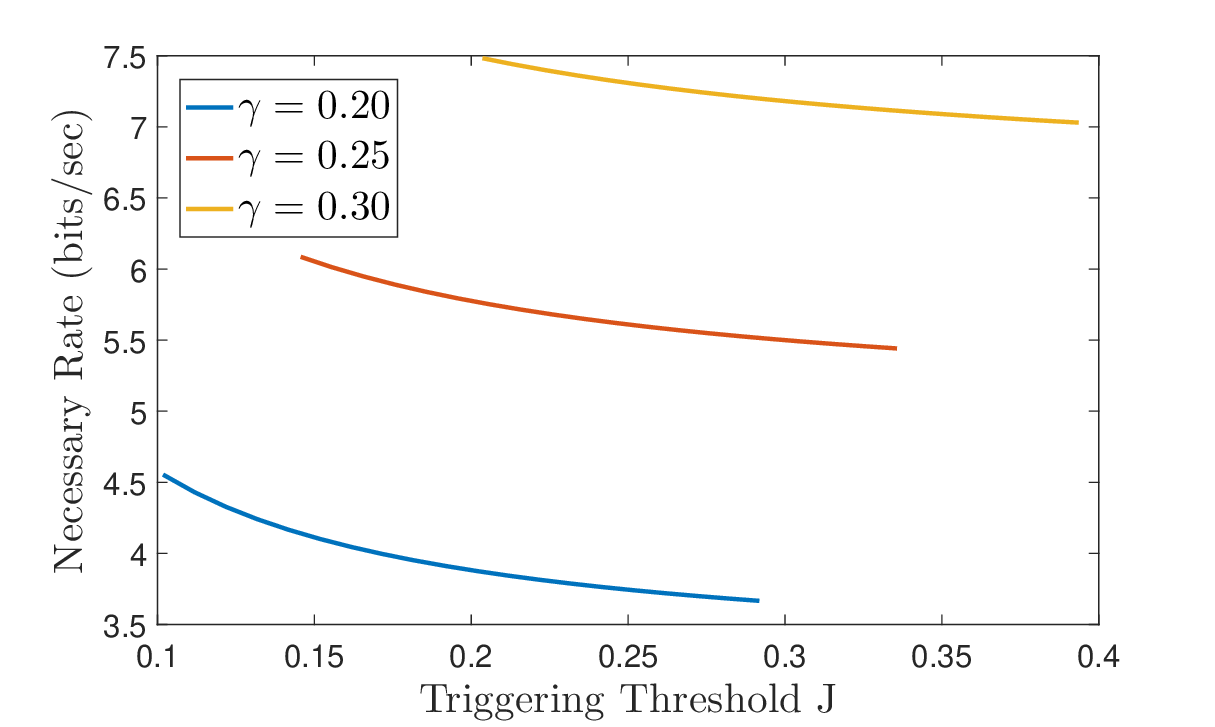}}
  \caption{(a) Sufficient and (b) necessary transmission rate versus triggering function $J$. Other simulation parameters are as follows: $A=5.5651$, $M=0.1$, $b=1.01$, and $\rho_0=0.1$.}
  \label{fig:graphs-J}
\end{figure}

\subsubsection{Design parameter $\rho_0$}\label{dempt53app}
Here we discuss the effects of the design parameter $\rho_0$ (which regulates the resolution of the quantization) on the sufficient rate~\eqref{inftranrate} via a numerical example.
According to~\eqref{Sufi!!221} and~\eqref{eq:jump-upp-suffpart}, any increase in $\rho_0$ results in a decrease in the packet size and an increase in $z(t_c^+)$ (the state estimation error after receiving the packet and updating the state). Consequently, a larger $\rho_0$ leads to higher triggering rates, cf.~\eqref{uppertrig1234}. 
\figref{fig:graphs-rs-rho0} illustrates the effects of the design parameter $\rho_0$ on the sufficient rate~\eqref{inftranrate}. In this numerical example, for small values of $\rho_0$, its effect on the packet size is dominant and the sufficient rate~\eqref{inftranrate} is a decreasing function of $\rho_0$. As $\rho_0$ keeps increasing, its effect on the  triggering rate starts becoming more dominant and the sufficient rate~\eqref{inftranrate} eventually becomes an increasing function of $\rho_0$.
\begin{figure}[ht]
  \centering
  \includegraphics[scale=0.40]{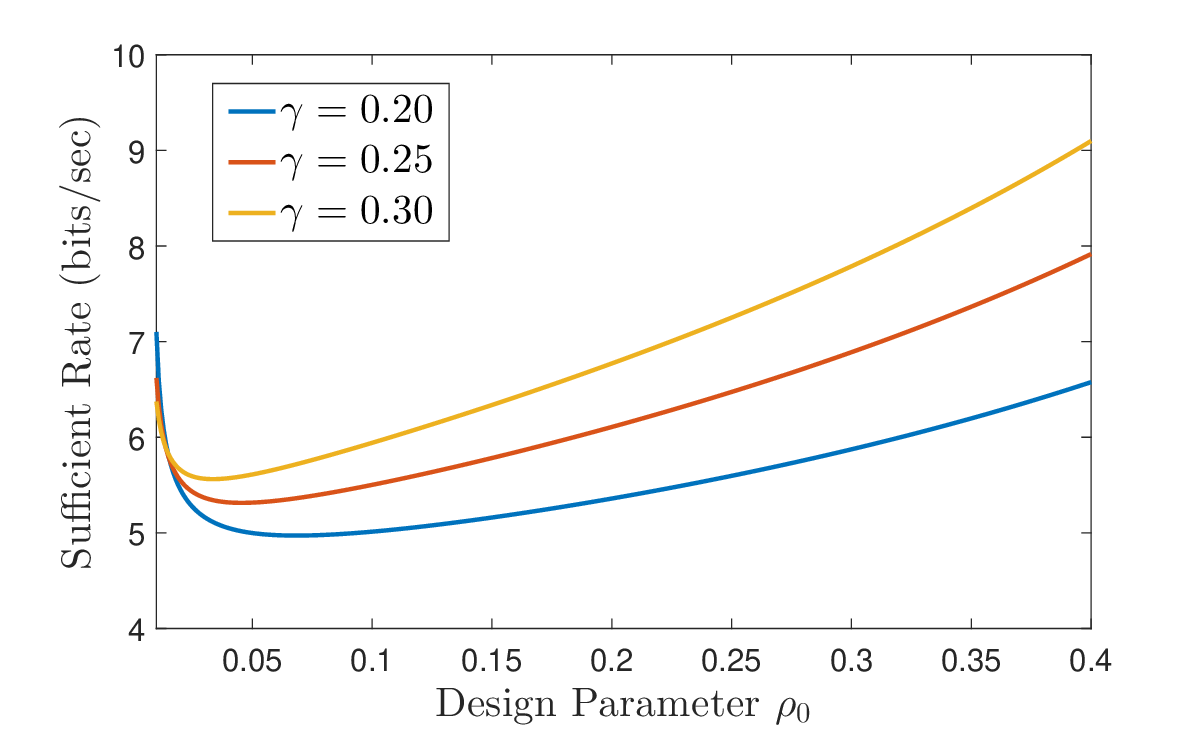} 
  \caption{Sufficient transmission rate versus design parameter $\rho_0$. Other simulation parameters are as follows: $A=2.50$, $M=0.5$, $b=1.01$, $\delta'=0.005$ seconds. The triggering threshold $J$ for the curves from top to bottom are 14.8882, 21.9065, and 31.1764, respectively.}
  \label{fig:graphs-rs-rho0}
\end{figure}


\subsubsection{Design parameter $b$}\label{sednb3312}
Here we discuss the effects of the design parameter $b$ on  the sufficient rate~\eqref{inftranrate} via a numerical example. The role of $b$ is explained as follows. Since the delay at the communication channel is upper bounded by $\gamma$ at the reception time $t_c$, the controller knows the triggering time $t_s$ belongs to the interval $[t_c-\gamma,t_c]$. Since $t_c$ is unknown at the time of transmission, as discussed in Section~\ref{sec:realn!}, to identify a unique interval (which the triggering time $t_s$ belongs to) that can be used as a common reference frame for the quantization of the transmission time for the sensor and the controller, we break the non-negative real number line into intervals of length $b\gamma$. This way, using $p(t_s)[2]$ and the fact that the controller knows $t_s \in [t_c-\gamma,t_c]$, the controller can identify an interval of length $b\gamma$ which $t_s$ belongs to, and the remaining bits of the packet are used to break this pre-specified interval further, cf.~\figref{encoder}. Clearly, to ensure such an interval can be determined with a single bit, we need to have $b >1$. As $b$ gets larger, the size of the pre-specified interval increases, and hence the sensor needs to put more bits in the packet to achieve a fixed resolution for the quantization of triggering time $t_s$. This is in agreement with~\figref{fig:graphs-rs-b} that illustrates our sufficient rate~\eqref{inftranrate} is an increasing function of the design parameter $b$. 

\begin{figure}[t]
  \centering
  \includegraphics[scale=0.40]{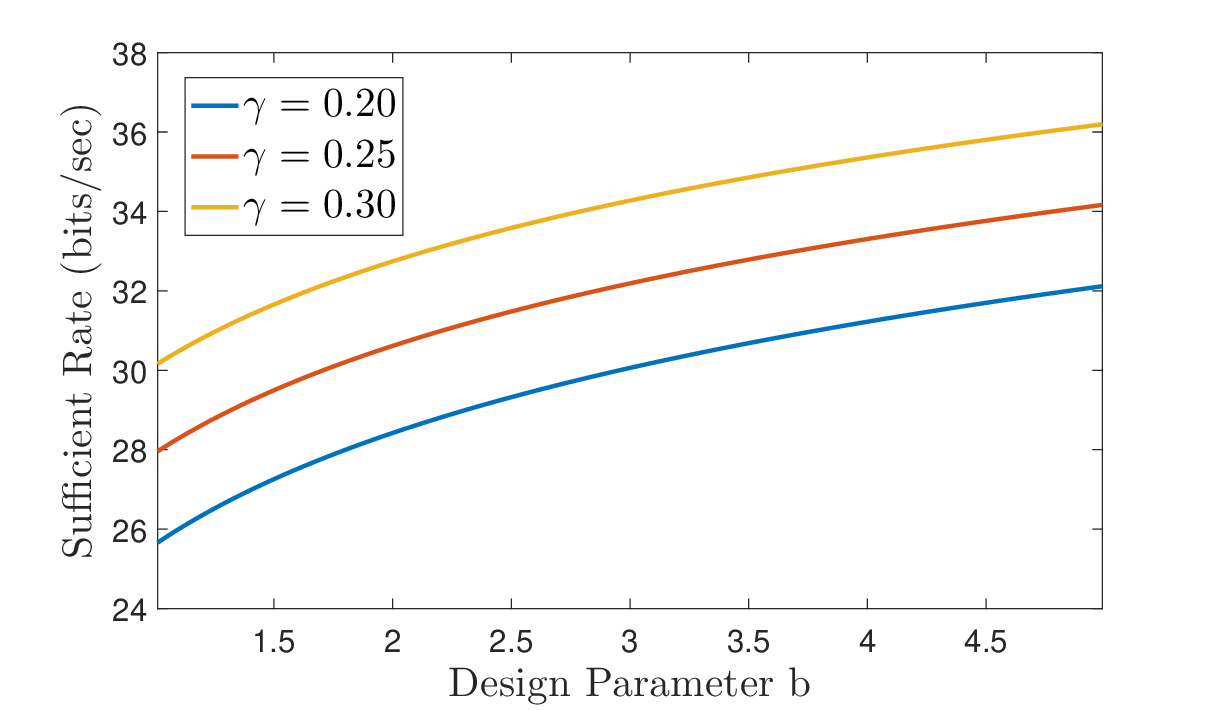} 
  \caption{Sufficient transmission rate versus design parameter $b$. Other simulation parameters are as follows: $A=5.5651$, $M=0.4$, $\rho_0=0.1$, and $J=(M/(A\rho_0))(e^{A\gamma}-1)+0.2$.}
  \label{fig:graphs-rs-b}
\end{figure}

\begin{remark}
In the case, $g(t_s)=1$ there is no need for $p(t_s)[2]$, and design parameter $b$, as described in~\figref{encoder}. In fact, for sufficiently small $\gamma$, the uncertainty about the value of the state at the controller is small, and the plant can be stabilized without breaking down the uncertainty set  $[t_c-\gamma,t_c]$ which the triggering time $t_s$ belongs to. 
\end{remark}

\begin{remark}
As discussed in Appendices~\ref{TTLSECr33} and~\ref{sednb3312}, larger $J$ and smaller $b$ lead to lower information transmission rates. Also, as discussed in Appendix~\ref{dempt53app}, $\rho_0$ has a different effect on the triggering rate and packet size. The investigation of the optimal choices for design parameters $\rho_0$, $J$, and $b$, as well as studying their effect on the trade-off between information transmission rate and control performances, is an interesting research venue.
\end{remark}

}}
\newpage
\section{Encoding and decoding algorithms}\label{sec:pccode}
In this section, we provide the pseudo-code descriptions of the encoding and decoding algorithms used in Section~\ref{sec:Stability} to derive our sufficient information transmission rate \footnote{{\blue{Implementation for the pseudo-codes can be found online at:

      \texttt{https://github.com/mkhojas/Event-Triggered-Firmware}.}
      }}.

\begin{algorithm}
\caption{Encoder}\label{alg:encoder}
\begin{algorithmic}
    \REQUIRE $t_s$, $b > 1$, $g(t_s)$, $\gamma$, $z(t_s)$
        \STATE $p(t_s)$ is an array of size $g(t_s)$ from~\eqref{efkfek!!!45hgree},  
        \STATE $p(t_s)[k] \in \{0,1\}$, $k=1, 2, 3, \textellipsis $ 
        \IF{$z(t_s) \geq 0$}
            \STATE $p(t_s)[1] \xleftarrow{} 0$
        \ELSE
            \STATE $p(t_s)[1] \xleftarrow{} 1$
        \ENDIF
        \STATE $i \xleftarrow{} \floor{\frac{t_s}{b\gamma}}$
        \STATE $[\alpha, \beta] \xleftarrow{} [ib\gamma, (i+1)b\gamma]$
        \STATE $i \xleftarrow{} 2$
        \WHILE{$i \leq g(t_s)$}
            \IF{$i = 2$}
                \STATE $p(t_s)[i] \xleftarrow{} \modulo \big( \floor{ \frac{t_s}{b\gamma} }, 2 \big)$
            \ELSE
                \IF{$t_s \in [\alpha, \frac{\alpha+\beta}{2}]$}
                    \STATE $p(t_s)[i] \xleftarrow{} 0$
                    \STATE $[\alpha, \beta] \xleftarrow{} [\alpha, \frac{\alpha+\beta}{2}]$
                \ELSE
                    \STATE $p(t_s)[i] \xleftarrow{} 1$
                    \STATE $[\alpha, \beta] \xleftarrow{} [\frac{\alpha+\beta}{2}, \beta]$
                \ENDIF
            \ENDIF
        \ENDWHILE
    \RETURN $p(t_s)$
\end{algorithmic}
\end{algorithm}

\begin{algorithm}[th]
\caption{Decoder}\label{alg:decoder}
\begin{algorithmic}
\REQUIRE $t_c$, $b > 1$, $\gamma$, $p(t_s)$, $J$, $A$
    \STATE $\bar{z}(t_c)$ is the estimate of $z(t_c)$ at the controller
    \STATE $q(t_s)$ is the estimate of $t_s$ after decoding the packet $p(t_s)$
    \STATE $g(t_s)$ is the length of $p(t_s)$ array
    \IF{$g(t_s) = 1$}
        \IF{$p(t_s)[1] = 0$}
            \STATE $\bar{z}(t_c) \xleftarrow{} J$
        \ELSE 
            \STATE $\bar{z}(t_c) \xleftarrow{} -J$
        \ENDIF
    \ELSE
        \STATE $i \xleftarrow{} \floor{\frac{(t_c-\gamma)}{b\gamma}}$
        \STATE $j \xleftarrow{} \floor{\frac{t_c}{b\gamma}}$
        
        \IF{$i=j$ or $\modulo(i, 2)=p(t_s)[1]$}
            \STATE $[\alpha, \beta] \xleftarrow{} [ib\gamma, (i+1)b\gamma]$
        \ELSE
            \STATE $[\alpha, \beta] \xleftarrow{} [jb\gamma, (j+1)b\gamma]$
        \ENDIF
        \STATE $i \xleftarrow{} 2$
        \WHILE{$i \leq g(t_s)$}
            \IF{$p(t_s)[i] = 0$}
                \STATE $[\alpha, \beta] \xleftarrow{} [\alpha, \frac{\alpha+\beta}{2}]$
            \ELSE 
                \STATE $[\alpha, \beta] \xleftarrow{} [\frac{\alpha+\beta}{2}, \beta]$
            \ENDIF
        \ENDWHILE
        \STATE $q(t_s) \xleftarrow{} \frac{\alpha+\beta}{2}$
        \IF{$p(t_s)[1] = 0$}
            \STATE $\bar{z}(t_c) \xleftarrow{} Je^{A(t_c-q(t_s))}$
        \ELSE 
            \STATE $\bar{z}(t_c) \xleftarrow{} -Je^{A(t_c-q(t_s))}$
        \ENDIF
    \ENDIF 
\RETURN $\bar{z}(t_c)$
\end{algorithmic}
\end{algorithm}

\end{document}